\documentclass[%
 reprint,
superscriptaddress,
 amsmath,amssymb,
 aps,
floatfix,longbibliography
]{revtex4-2}

\usepackage{graphicx}
\usepackage{dcolumn}
\usepackage{bm}
\usepackage{siunitx}
\usepackage{float}
\usepackage{natbib}
\usepackage{appendix}
\usepackage{ulem}
\usepackage[pagewise]{lineno}
\usepackage{xcolor}

\newcommand{\add}[1]{\textcolor{black}{#1}}

\newcommand{\Add}[1]{\textcolor{black}{#1}}
\newcommand{\Addd}[1]{\textcolor{black}{#1}}
\newcommand{\Adddd}[1]{\textcolor{black}{#1}}

\begin{document}

\preprint{APS/123-QED}

\title{Why epithelial cells collectively move against a traveling signal wave} 

\author{Tatsuya Fukuyama}
\affiliation{Department of Chemical Engineering, Kyoto University, Kyoto 615-8246, Japan}

\author{Hiroyuki Ebata}
\affiliation{Department of Earth and Space Science, Graduate School of Science, Osaka University, Osaka 560-0043, Japan}

\author{Akihisa Yamamoto}
\affiliation{Department of Mechanical Systems Engineering, Tokyo Metropolitan University, Tokyo 192-0397 Japan}

\author{Ryo Ienaga}
\affiliation{Department of Chemical Engineering, Kyoto University, Kyoto 615-8246, Japan}

\author{Yohei Kondo}
\affiliation{NU-TLiMP, Nagoya University, Aichi 466-8550, Japan}

\author{Motomu Tanaka}
\affiliation{Center for Integrative Medicine and Physics, Institute for Advanced Study, Kyoto University, Kyoto 606-8501, Japan}\affiliation{Physical Chemistry of Biosystems, Institute of Physical Chemistry, Heidelberg University, D69120 Heidelberg, Germany}

\author{Satoru Kidoaki}
\affiliation{Institute for Materials Chemistry and Engineering, Kyushu University, Fukuoka 819-0395, Japan}
\author{Kazuhiro Aoki}
\affiliation{Laboratory of Cell Cycle Regulation, Graduate School of Biostudies, Kyoto University, Kyoto 606-8501, Japan}\affiliation{Center for Living Systems Information Science (CeLiSIS), Graduate School of Biostudies, Kyoto University, Kyoto 606-8501, Japan}

\author{Yusuke T. Maeda}\email{maeda@cheme.kyoto-u.ac.jp}\affiliation{Department of Chemical Engineering, Kyoto University, Kyoto 615-8246, Japan}


\date{\today}

\begin{abstract}
The response of cell populations to external stimuli plays a central role in biological mechanical processes such as epithelial wound healing and developmental morphogenesis. Wave-like propagation of a signal of ERK MAP kinase has been shown to direct collective migration in one direction; however, the mechanism based on continuum mechanics under a traveling wave is not fully understood. To elucidate how the traveling wave of the ERK kinase signal directs collective migration, we constructed the mechanical model of the epithelial cell monolayer by considering the signal-dependent coordination of contractile stress and cellular orientation. The proposed model was studied by using an optogenetically-controlled cell system where we found that local signal activation induces changes in cell density and orientation with the direction of propagation. The net motion of the cell population occurred relative to the wave, and the migration velocity showed a maximum in resonance with the velocity of the ERK signal wave. The presented mechanical model was further validated in \textit{in vitro} wound healing process.
\end{abstract}

\maketitle

\section{Introduction}
\Adddd{Collective migration refers to the cohesive motion of cells as a group, exhibiting a high degree of correlation in migration direction \cite{Friedl2009}. The cells align their direction of movement through cell adhesion to one another, thereby enabling the group of cells to move in a coordinated fashion. These collective cell migrations play essential roles in tissue remodeling processes \textit{in vivo}, as evidenced by the observation of embryonic development, wound healing of epithelial tissues, and cancer cell invasion \cite{Friedl2009, MATSUBAYASHI2004, AOKI2013, Hiratsuka2015, Aoki2017}.
\textit{In vitro} wound healing assay with cultured epithelial cells such as MDCK cells has been widely used as a model system for collective migration. Recent studies reveal that the direction of collective migration during \textit{in vitro} wound healing is regulated by intracellular signaling pathways \cite{weijer2009collective}. The extracellular signal-regulated kinase (ERK) pathway is a representative model in this regard \cite{MATSUBAYASHI2004}; the ERK MAP kinase is an intracellular signal transduction protein known to regulate cell proliferation, differentiation, migration, and apoptosis \cite{seger1995mapk}. During collective migration of epithelial monolayer cells, the ERK activity transiently increases in cells in the vicinity of the injured region. As ERK activity is upregulated in the cells, the ERK activation causes EGFR-ligand cleavage in those cells and subsequent the activation of EGFR-Ras-ERK pathway in the neighboring cells, leading to the propagation of ERK activation as a trigger wave through the cell monolayer \cite{AOKI2013,Aoki2017}. As the wave-like propagation of an ERK activity passes through, the direction of collective migration is guided in the direction opposite to that of the ERK wave \cite{Aoki2017, londono2014, Hiratsuka2015}. This rectification effect would be important for detecting and moving toward a wound to fill the gap in the cell population.}

How can the propagation of signaling activity across cell populations, such as the ERK activation waves, affect the collective migration of cells? A physical process that achieves a finite net motion is needed when the signal increases and decreases in one cycle as its wave passes. This requirement is because a spatially symmetric wave does not result in the net motion of cells as the forces generated by the wave are equal in magnitude but opposite in sign. One possible way to address this question is to consider the broken time-reversal symmetry that results from the multiplication of two physical processes \cite{purcell1977, najafi2004,lozano2019}: if the forces driving cell migration involve more than two processes, time-reversal symmetry can be broken, resulting in net motion, even at force-balance condition \cite{kumar2008, qiu2014, lauga2011, leoni2017, mai2020}. However, the existence of these multiplicative effects in collective migration remains unclear. Determining the relationship between directed motion and a traveling signal wave is an important challenge. 

Cell biophysics has contributed greatly to revealing the mechanical model for collective dynamics in a cell monolayer \cite{garcia2015physics, blanch2018turbulent, mueller2019emergence, lin2021energetics}. Although individual cells move randomly in the low-density condition, when their density increases, the polarity axis of the cells is oriented by adhesion with neighboring cells, resulting in aligned directions of movement and emergent collective migration. Thus, there is growing interest in elucidating the physical principles of complex tissue formation by characterizing the collective dynamics of cell populations \cite{perez2021mechanical,brandstatter2023curvature}. Numerical models of epithelial collective migration have shown that the orientation dynamics that align polarity and local cell density changes are coupled to produce spatiotemporally ordered structures \cite{honda2001vertex, fletcher2014vertex, barton2017active, lin2018dynamic}. In the context of wound healing, collective migration guided by ERK signal waves is closely related to the self-repair of epithelial structure, and elucidating the complex processes of epithelial tissue from the perspective of mechanics of collective migration directed by the wave of chemical signaling is important.

In this study, we investigate the mechanics of collective migration along a traveling wave of ERK MAP kinase. We show that by considering the propagation of ERK activation, which regulates both local cell contraction and the cell orientation field in a cell monolayer, directed collective migration can occur in the direction opposite to the ERK wave. Through experimental investigation, we explain key aspects of directed collective migration, such as the non-monotonic increase in migration speed with respect to the activation level and the speed of the ERK wave. Our results indicate that the traveling wave rectifies the direction of collective migration due to a mechanical effect: the interplay of signal-dependent contraction and anisotropic friction.

\add{\section{Summary of theoretical model analysis}}

\add{To elucidate the mechanisms underlying collective cell migration against a propagating ERK signal wave, we developed a continuum mechanical model that describes an epithelial cell monolayer as a viscoelastic fluid coupled to a traveling biochemical signal. In this framework, each cell's contractile force is regulated by the local activity of the ERK signal, which propagates across the monolayer as a wave. The model incorporates spatial variations in cell density, velocity as well as the orientation of cells. The ERK signal is mathematically treated as a traveling Gaussian profile, and the force balance equation for the monolayer is derived under low Reynolds number conditions.\\
To connect the complexity of intracellular signaling to population-level mechanics, we coarse-grain the ERK/MAPK cascade by assuming that ERK activity predominantly regulates both myosin contractility and cell orientation, as supported by experiments (Fig. 2 and Fig. S3) and previous studies \cite{AOKI2013,Aoki2017}. Other signaling intermediates are assumed to equilibrate rapidly and are incorporated into effective parameters within the model. This minimal description is designed to capture the physical mechanism by which a traveling ERK wave dynamically biases cellular mechanical properties to break time-reversal symmetry and directs collective cell migration.\\
The theoretical model demonstrates that collective migration arises from the interplay between ERK-dependent contractile force generation and the anisotropy of friction within the cell monolayer. By expanding the governing equations in terms of the small ERK signal amplitude, we show that the net velocity of the cell monolayer at steady state is proportional to the product of the ERK signal wave speed, the degree of anisotropy in friction, the degree of density change. Notably, the model predicts that the direction of net cell migration is opposite to that of the propagating ERK wave, consistent with experimental observations shown later (Figure 3).\\
Furthermore, the model can be extended to account for the viscoelastic nature of the cell monolayer, introducing a time-dependent relaxation term that captures the migration speed dependent on the speed of ERK wave propagation. This reveals that at low ERK wave speeds, the migration velocity increases linearly with wave speed; however, at higher wave speeds, viscoelastic relaxation limits the migration velocity, leading to a plateau and eventual decline (Figure 3). The framework was also generalized to model wound healing, where collective migration combines the effect of ERK-driven movement with the advance of the tissue boundary, yielding a quantitative description of observed healing dynamics (Figures 4 and 5).}

\section{Theoretical model of collective migration under signaling wave}

\begin{figure*}[tb]
\begin{center}
\includegraphics[width=15cm]{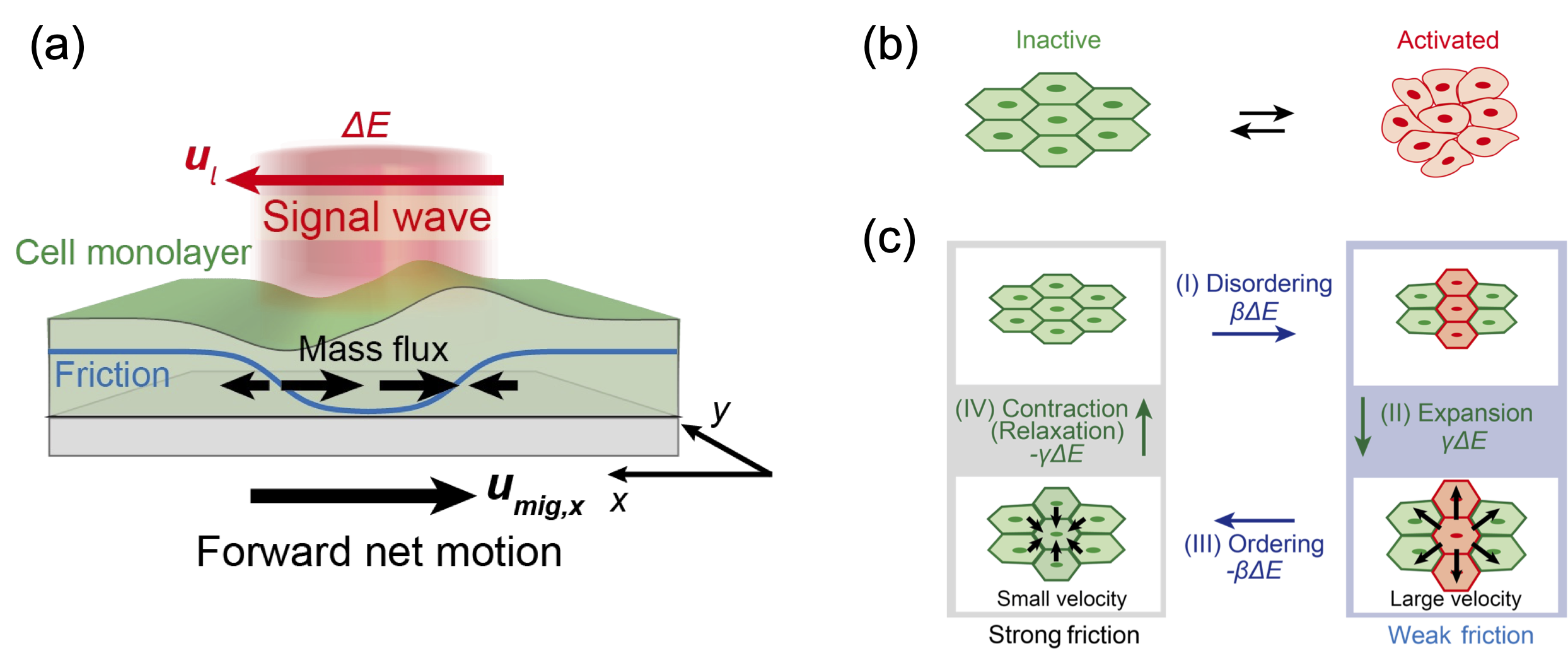}
 \caption{\textbf{Directed collective migration under a traveling signal wave.} (a) Schematic illustration of directed collective migration. The green color indicates a cell monolayer, the height of which represents the change in cell density. (b) Schematic of cell population activated by a signal protein. (c) Schematic illustration of the proposed multiplication of the active contraction and the cellular orientation. The activated signal reduces the anisotropic friction (I: disordering). Signal activation, in turn, causes a cell to be pulled by the surrounding cells (I\hspace{-.1em}I: expansion). As the signal wave goes through the cell monolayer, the activated cell restores the frictional force (I\hspace{-.1em}I\hspace{-.1em}I: ordering) and then it relaxes the pulling force (I\hspace{-.1em}V: contraction). Four distinct stages break the time-reversal symmetry in motion. \add{The motion shown in (a)-(c) corresponds to the case of $\epsilon < 0$, $\beta <0$, and $\gamma<0$.} }\label{fig1}
\end{center}\end{figure*}

To study how signaling waves control collective migration, we used continuum mechanics. The cell population regulated by intracellular protein signaling can be described as a continuum of two-dimensional coordinates $\bm{x} = (x, y)$ (Fig. \ref{fig1}(a)). This theoretical model aims to demonstrate the mechanism controlling collective migration (forward net motion) along a wave of protein signals as it travels through a cell monolayer. Although collective migration is regulated by complex signaling pathways in real cells, it is simplified here to assume that the cell population is activated with a high ERK signal level as it receives ERK kinase signal but this active state gradually decays and returns to the basal state with a low ERK signal level (Fig. \ref{fig1}(b)). The basal level of the protein signal is given by $E_0$ and takes the same constant value at infinity regardless of the direction. The ERK signal intensity is defined by $\Delta E(\bm{x},\Adddd{t})$ = $E(\bm{x},t) - E_0$, and we assume that the ERK signal is spread by the point source propagating at a constant velocity $\bm{u_l} = (u_l, 0)$. We consider the point source of the ERK signal to have a Gaussian form $\Delta E_0 \exp[-(\bm{x}-\bm{u_l}t)^2/2a^2]$ with the width of $a$. \add{The Gaussian distribution reflects the diffusion-like propagation of signal activity across the cell monolayer. This modeling choice is based on experimental observations showing that ERK activation spreads to neighboring cells \cite{AOKI2013,Aoki2017}. Given that diffusion from a localized source naturally gives rise to a Gaussian profile, we adopted this form for analytical convenience while capturing the essential biological mechanism of signal propagation.} The peak size of the ERK signal is $\Delta E_0$, and in the following analysis, we assume that ERK signal changes follow this distribution function $\Delta E(\bm{x},t)$ in two-dimensional space \cite{supplement}. 

Collective migration is driven by internal forces while maintaining force-free conditions \cite{wang2011, tanimoto2014, ebata2020, tarama2018,braun1,mae1}. The cell population adheres to the substrate and balances the contractile stresses generated by the cells and frictional stress. We considered that both the contractile force and frictional stresses depend on the magnitude of signal activation. The force balance is described as follows:
\begin{equation}\label{stokeseq}
\rho \frac{\partial \bm{u}}{\partial t} = - \bm{\zeta} \bm{u} + \bm{\nabla} \cdot \bm{\sigma} 
\end{equation}
where $\rho = \rho(\bm{x}, t)$ denotes the local cell density, $\bm{u}(\bm{x}) = (u_x(\bm{x}), u_y(\bm{x}))$ denotes the velocity field of the cell population, and $\bm{\zeta}$ denotes the frictional coefficient matrix for cell migration per unit area. Cells adhere to each other via cell-cell adhesion junctions, where $\bm{\sigma}$ represents the internal stress \cite{banerjee2015, yabunaka2017}. For an epithelial cell monolayer typically moving at low velocity, the inertia term is negligible, and Eq. \eqref{stokeseq} is rewritten as
\begin{equation}\label{forcenalance}
\bm{\nabla} \cdot \bm{\sigma} = \bm{\zeta}\bm{u}.
\end{equation}
We defined the internal stress as
\begin{equation}\label{stress}
\bm{\sigma} = -\Pi \bm{I} + k c \bm{I}
\end{equation}
where $\bm{I}$ denotes the identity matrix, the first term denotes the passive stress caused by the pressure $\Pi$, and the second term denotes the ERK signal-dependent stress generated by the molecular motor protein (protein concentration $c$) with a constant $k$ \cite{banerjee2015}. \add{We assumed the contractile stress associated with ERK activation to be isotropic at the scale of the monolayer and thus represented by the identity matrix. This is consistent with the assumption that myosin-II generates uniform cortical contractility in a cell but the anisotropic effects arise from signaling gradients \cite{Aoki2017}.} However, there is no net motion under changes in density $\rho$ when the friction coefficient $\bm{\zeta}$ does not change with signal activation \cite{braun1, mae1}. This is because the net internal stress (i.e., integral of $\nabla \Pi $ over the cell population) must be zero under force-free conditions. The active stress induced by the molecular motors is insufficient to rectify the directionality of motion relative to the traveling signal wave.

The key concept that defines directed net motion is the multiplication of the frictional force and active contractile force regulated by the protein signal (Fig. \ref{fig1}(c)). Past studies have hypothesized an ERK signaling dependence on adhesion with respect to friction \cite{Aoki2017,asakura2021hierarchical}, but the specific physical process remains unclear. Instead, we assume that the orientation of the MDCK cell population determines the anisotropy of the friction coefficient and that ERK signaling regulates such anisotropic orientation, which allows us to give a more specific model of the signal dependence of the frictional force.

Cells have an orientation within the cell population, which determines the ease of movement and affects the direction of cell migration \cite{saw2017, kawaguchi2017, doostmohammadi2018}. The orientation of the cell monolayer is expressed by the director $\bm{n}(\bm{x}) = (n_x, n_y) = (\cos\theta, \sin\theta)$, with an orientation angle of $\theta(\bm{x})$. If this orientation is changed by the signaling activity, the anisotropic mobility may be altered depending on the orientation field and spatially biased by $\bm{n}$. 

The tensor of the orientation field is defined as $\bm{Q} = S(\bm{n}\bm{n} - \frac{1}{2}\bm{I})$ where the scalar order parameter $S = \sqrt{\langle \cos2\theta \rangle^2 +\langle \sin 2\theta \rangle^2}$ is determined from the average $\theta$ within the coarse-grained region. We assume that $S$ changes in a signal-dependent manner (i.e., $S = S(E)$). The anisotropic friction $\bm{\zeta}$ is given as follows:
\Adddd{\begin{eqnarray}\label{friction}
\bm{\zeta} &=& \begin{bmatrix}
   \zeta_{xx} & \zeta_{xy} \\
   \zeta_{yx} & \zeta_{yy}
\end{bmatrix} = \zeta_0(\bm{I} - \epsilon \bm{Q}) \nonumber \\ 
&=& \zeta_0 \begin{bmatrix}
   1- (\epsilon S/2)\cos2\theta & (\epsilon S/2)\sin2\theta \\
   (\epsilon S/2)\sin2\theta & 1+(\epsilon S/2)\cos2\theta 
   \end{bmatrix}
\end{eqnarray}}
where $\epsilon$ is the scalar friction parameter representing orientational anisotropy \cite{kawaguchi2017}. \Addd{Negative (or positive) $\epsilon$ means that the coefficient of friction for cell migration perpendicular to the orientation angle increases (or decreases).} Eq. (4) then gives the viscous resistance force as follows:
\Adddd{\begin{eqnarray}\label{friction2}
\bm{\zeta}\bm{u} &=& \begin{bmatrix}
   \zeta_{xx} u_{x}+ \zeta_{xy}u_y \\
   \zeta_{yx} u_{x}+ \zeta_{yy}u_y 
\end{bmatrix}
\end{eqnarray}}
\Adddd{We consider a situation in which the cell population migrates in $x$-axis direction along which the ERK wave propagates.} By substituting Eqs. (3) and \eqref{friction2} into Eq. \eqref{forcenalance}, the force-balance relation becomes
\begin{equation}\label{balance2x}
\zeta_{xx} u_x + \zeta_{xy}u_y = \frac{\partial}{\partial x}\Bigl( - \Pi + kc \Bigr),
\end{equation} and
\begin{equation}\label{balance2y}
\zeta_{xy} u_x + \zeta_{yy}u_y = \frac{\partial}{\partial y}\Bigl(- \Pi + kc \Bigr),
\end{equation}
where we use $\zeta_{xy}=\zeta_{yx}$. Eqs. (6) and (7) lead 
\begin{equation}\label{balance2all}
\frac{\partial}{\partial y}\bigl(\zeta_{xx}u_x+\zeta_{xy}u_y\bigr) - \frac{\partial}{\partial x}\bigl(\zeta_{yx} u_x + \zeta_{yy}u_y \bigr) =0.
\end{equation}

Next, we analyze the mass conservation with a continuum equation. Given that the signal continues to activate the cell body until the wave passes, contractile stress is induced and becomes stronger at the rear of the wave \cite{saraswathibhatla2022, yang2018}. The resultant gradient of the contractile stress changes the local cell density $\rho$, which follows the continuity equation
\begin{equation}\label{continuum}
\frac{\partial \rho}{\partial t} + \bm{\nabla} \cdot (\rho \bm{u}) = 0.
\end{equation}
\Add{The characteristic time for cell motility is typically estimated as the time required for a cell to traverse a distance approximately equal to its size at a velocity of 15 - \SI{20}{\micro\meter\per\hour}, roughly 1 h. Given that cell division of typical epithelial cells occurs approximately once every 12 h, we consider these time scales to be significantly different. Therefore, we assume that the impact of cell division is negligible}. 

\Adddd{To find $\bm{u}(\bm{x})$ that satisfies Eqs. \eqref{balance2all} and \eqref{continuum}, we use the moving frame coordinate (\Adddd{$\bm{x'}=\bm{x}-\bm{u}_lt$}, $\frac{\partial}{\partial t} = \Adddd{-} u_l\frac{\partial}{\partial x'}$), where the protein signal travels at a constant velocity $\bm{u}_l = (u_l, 0)$, and note that $u_x$, $u_y$, and $\rho$ have solutions that depend only on ($x'$, $y$).
This coordinate transformation rewrites Eqs. \eqref{balance2all} and \eqref{continuum} as
\begin{equation}\label{balance3}
\frac{\partial}{\partial y}\bigl(\zeta_{xx}u_x+\zeta_{xy}u_y\bigr) - \frac{\partial}{\partial x'}\bigl(\zeta_{yx} u_x + \zeta_{yy}u_y \bigr) =0,
\end{equation}
and
\begin{equation}\label{continuum2}
-u_l \frac{\partial \rho}{\partial x'} + \frac{\partial }{\partial x'}(\rho u_x)+\frac{\partial }{\partial y}(\rho u_y)=0. 
\end{equation}}

\begin{figure*}[tb]
\begin{center}
\includegraphics[width=11cm]{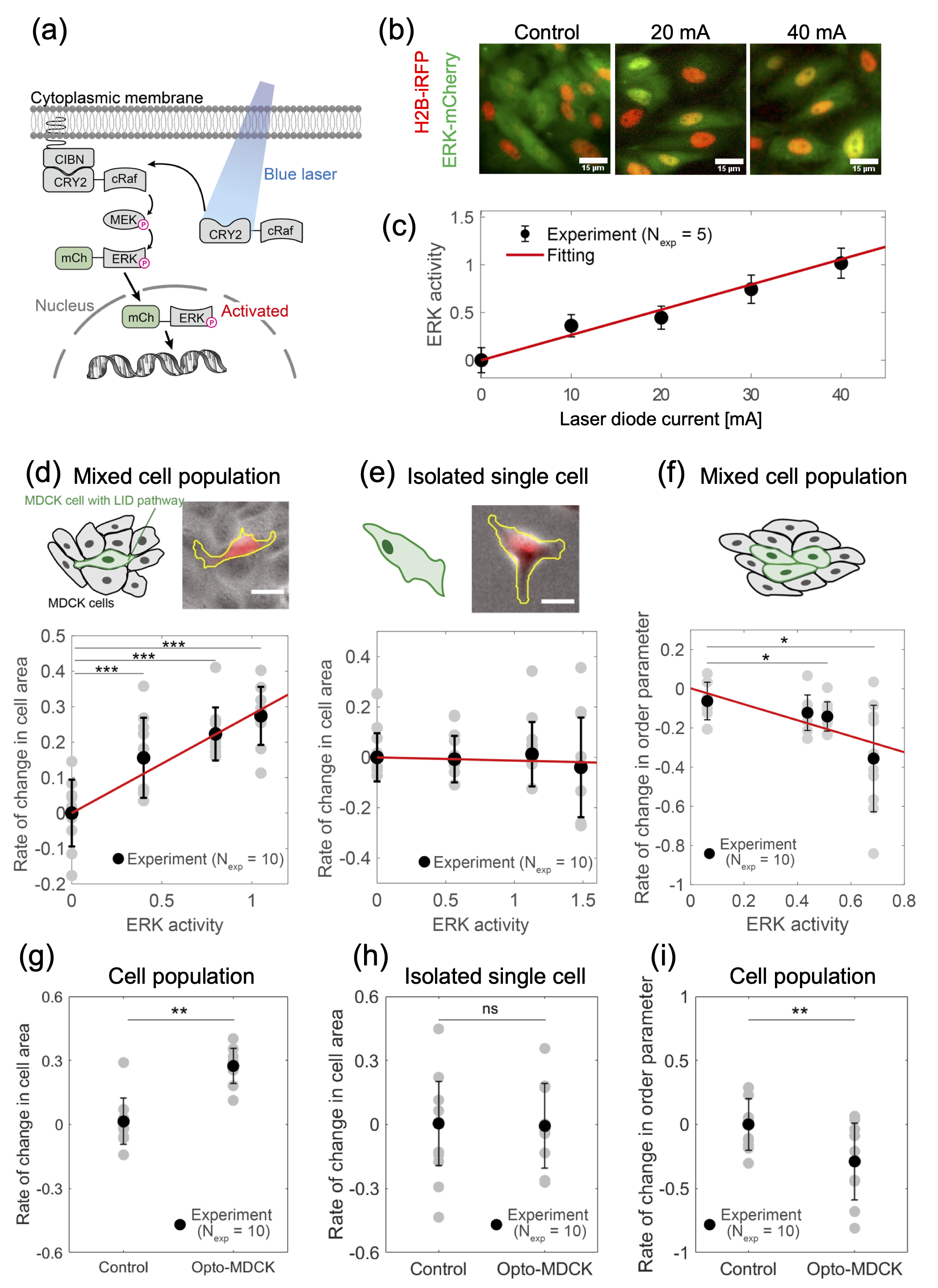}
\caption{\textbf{Controlling epithelial cells with optogenetic ERK signal activation}. (a) Optogenetic control of ERK activity through light-induced dimerization. (b) Fluorescent imaging analysis for quantifying ERK activity, $\Delta E$. In the absence of light, ERK-mCherry is found in the cytoplasm. ERK signaling induces the nuclear translocation of ERK-mCherry (denoted by H2B-iRFP in red) through transduction. The extent of nuclear migration is directly proportional to the degree of substrate phosphorylation by the ERK MAP kinase. ERK activity can be regulated quantitatively by adjusting light intensity. Scale bars: \SI{15}{\micro\meter}. (c) Linear calibration plot of ERK activity for optogenetic light intensity. The experimental data is represented by black dots. The red line represents the linear curve with the slope $g$=0.022 with 95\% confidential interval (CI) of 0.011 to 0.029. The numbers of technical replicates is $n=1$ and the number of independent experiments is $N_{exp}=5$. (d) Changes in the area of ERK-activated cells in the mixed cell population (bottom). The red line is the fitting linear curve with the slope $g$=0.25 (95\% CI is 0.18 to 0.33). Schematic illustration of light-inducible dimerization (LID) pathway in the mixed cell population (top). A cell with LID pathways was outlined with the yellow line, and the nucleus is shown in red. (e) Changes in cell area in isolated single cells (bottom). The red line is the fitting linear curve with the slope $g$=-0.01 (95\% CI is -0.11 to 0.09). Schematic of single cells activated through the LID pathway (top). (f) Changes in orientation order in ERK-activated cells in the mixed cell population (bottom). The red line is the fitting linear curve with the slope $g$=-0.44 (95\% CI is -0.71 to -0.17). Data were analyzed using the Mann-Whitney U-test, *** $p < 0.001$, * $p < 0.05$. In (d)-(f), the number of technical replicates is $n=2$ for (d) and (e) or $n=3$ for (f), and the number of independent experiments is $N_{exp}$ = 10. The black circle indicates the mean value and the error bar is the SD. \add{We note that the standard deviations and resulting confidence intervals for area and orientational order are relatively large in our measurements. This variability is attributable not only to differences in ERK activation but also to heterogeneity in the initial state of the cell monolayer. Although the overall trends were consistent, the extent of these changes varied depending on the initial conditions. To confirm that the light‑induced effects in panels (d)–(f) result specifically from optogenetic ERK activation, we compared wild‑type MDCK cells (which lack the LID pathway) with optogenetic MDCK cells. Both cell types were exposed to blue light (g: the cell population, h: isolated cells, i: the cell population).  In wild‑type controls, no significant changes were observed upon light stimulation (laser diode current: 40 mA)in any of these assays. Statistical comparisons were made using the Mann–Whitney U-test, ** $p < 0.01$.}}\label{fig2}
\end{center}
\end{figure*}

\Adddd{To solve these equations for perturbations of ERK signal change, we expand the velocity of motion $u_i$ ($i=x, y$) with peak size of the ERK signal $\Delta E_0$,
\begin{equation}
u_i = u_{i1}+ u_{i2}+\cdots
\end{equation}
where 1 and 2 represent the 1st order and 2nd order of perturbation $\Delta E_0$, respectively. The cell population does not move without the ERK signal, i.e., $u_{i0}=0$. We also expand $\bm{\zeta}$ and $\rho$ with $\Delta E_0$ and consider up to the first order written by
\begin{eqnarray}\label{perturbation}
\zeta_{xx} &=& \zeta_{xx0} + \zeta_{xx1} + \cdots \nonumber \\ &=& \zeta_{xx0} (1 + \beta \Delta E) + \cdots \\
\zeta_{yy} &=& \zeta_{yy0} + \zeta_{yy1} + \cdots \nonumber \\ &=& \zeta_{xx0} (1 - \beta \Delta E) + \cdots \\
\zeta_{xy} &=& \zeta_{xy0} + \zeta_{xy1} + \cdots \nonumber\\  &=& \zeta_{xy0} + \zeta_{xx0}\beta' \Delta E + \cdots \\
\rho &=& \rho_0 + \rho_1 + \cdots \nonumber \\ &=& \rho_0(1-\gamma a^2 \nabla^2(\Delta E))+ \cdots 
\end{eqnarray}
where $\beta = \frac{1}{\zeta_{xx}}\frac{\partial \zeta_{xx}}{\partial E}$, $\beta' = \frac{1}{\zeta_{xx}}\frac{\partial \zeta_{xy}}{\partial E}$, $\gamma = \frac{\partial \ln\rho}{\partial E}$, and $a$ is the width of the ERK signal distribution. We also assume that $S$ is a minute parameter as small as $\Delta E$. In the density expansion presented in Eq. (16), we assumed the simplest terms that satisfy the mass conservation law and spatial symmetry for any $\Delta E$. The validity of $\beta$ parameter will be verified in the experiment by the fact that cells tend to move in the oriented direction shown later in Fig. \ref{fig2}.
We can write down the expression consisting of the first order terms of $\Delta E_0$ in Eq. (10), by considering $u_{x0}=u_{y0}=0$,
\begin{equation}\label{balance4}
\frac{\partial}{\partial y}\bigl(\zeta_{xx0}u_{x1}+\zeta_{xy0}u_{y1}\bigr) - \frac{\partial}{\partial x'}\bigl(\zeta_{xy0} u_{x1} + \zeta_{yy0}u_{y1} \bigr) =0.
\end{equation}
Because $\zeta_{xy0}$ is $\mathcal{O}(S)$ but $\zeta_{xx0}=\zeta_{yy0}=\zeta_0$ are $\mathcal{O}(1)$, $\zeta_{xy0}u_{y1}$ and $\zeta_{xy0} u_{x1}$ can be negligible. By this approximation, Eq. \eqref{balance4} becomes vortex-free relation $\frac{\partial u_{x1}}{\partial y} - \frac{\partial u_{y1}}{\partial x'}=0$.}

Furthermore, we find the relation that is also valid for the density $\rho$ in the first order of $\Delta E_0$ from Eq. \eqref{continuum2}. 
\begin{equation}\label{continuum4}
-u_l \frac{\partial \rho_1}{\partial x'} + \rho_0 \frac{\partial u_{x1}}{\partial x'}+\rho_0\frac{\partial u_{y1}}{\partial y}=0. 
\end{equation}
Using $\rho_1=-\rho_0\gamma a^2 \nabla^2 \Delta E$, Eq. \eqref{continuum4} is rewritten by
\begin{equation}\label{continuum5}
-u_l \gamma a^2 \frac{\partial}{\partial x'} \nabla^2 \Delta E=  \frac{\partial u_{x1}}{\partial x'}+\frac{\partial u_{y1}}{\partial y}. 
\end{equation}
The solution of $u_{x1}$ that satisfies Eqs. \eqref{balance4} and \eqref{continuum5} is
\begin{equation}\label{vel1st}
u_{x1} = - u_l\gamma a^2 \frac{\partial^2\Delta E}{\partial x'^2}.
\end{equation}
If $\Delta E= \partial\Delta E/\partial x'=0$ at infinity ($x, y \rightarrow \infty$), the spatial average of $\partial^2\Delta E/\partial x'^2$ is also zero. We thus find that spatial average of $u_{x1}$ is also zero, meaning that net motion is not realized within the first order. Therefore, to explain collective motion under signal waves, we need to solve for $\bm{u}_{2}$.

We write the second-order term of the ERK signal change in Eq. (10). Since $\zeta_{xy0}$ is small in $\mathcal{O}(S)$, we can ignore the terms that include $\zeta_{xy0}$ as coefficient: 
\begin{equation}\label{balance5}
\frac{\partial}{\partial y}\bigl(\zeta_{xx1}u_{x1}+\zeta_{xx0}u_{x2}+\zeta_{xy1}u_{y1}\bigr) - \frac{\partial}{\partial x'}\bigl(\zeta_{xy1} u_{x1} + \zeta_{yy1}u_{y1} \bigr) =0.
\end{equation}
We consider that there is no net motion in the direction perpendicular to the ERK wave (see detailed calculation in \cite{supplement}). Eqs. (13)-(16) give $\zeta_{xx1}/\zeta_{xx0}=\beta\Delta E$, $\zeta_{yy1}/\zeta_{xx0}=-\beta\Delta E$, and $\zeta_{xy1}/\zeta_{xx0}=\beta'\Delta E$. By substituting these coefficient relationships into Eq. \eqref{balance5} and integrating in the $y$ direction, $u_{x2}$ is determined by
\begin{eqnarray}\label{vel2nd1}
& &u_{x2}= -u_{x1}\beta\Delta E - u_{y1}\beta' \Delta E \nonumber \\& &-\beta\frac{\partial}{\partial x'}\int_{-\infty}^{y} dy u_{y1}\Delta E + \beta'\frac{\partial}{\partial x'}\int_{-\infty}^{y} dy u_{x1}\Delta E.
\end{eqnarray}
We are interested in the spatial average in the $x$ direction, $\langle u_{x2} \rangle_x$. From the symmetric distribution of $E$, $u_{x1}$ and $\Delta E$ are even functions in $x$ and $y$ while $u_{y1}$ is an odd function. Considering that $\Delta E$ and $u_{y1}$ are zero at the far end, only the first term on the right side of Eq.\eqref{vel2nd1} contributes as
\begin{equation}\label{vel2nd2}
\frac{1}{\sqrt{\pi}a} \int_{-\infty}^{\infty} dx' u_{x2} \equiv \langle u_{x2} \rangle_{x'} =  - \langle u_{x1}\beta\Delta E \rangle
\end{equation}
(see detailed calculation in \cite{supplement}). We recall that $u_{x1}$ is derived as Eq. \eqref{vel1st} from the continuum equation. By performing a partial integral, we obtain
\begin{eqnarray}\label{vel2nd3}
\langle u_{x2} \rangle_{x'} &=& u_l\beta\gamma \frac{a}{\sqrt{\pi}} \int_{-\infty}^{\infty} dx' \Delta E \frac{\partial^2\Delta E}{\partial x'^2} \nonumber \\
&=& -u_l\beta\gamma a^2 \Big\langle\Bigl(\frac{\partial\Delta E}{\partial x'}\Bigr)^2\Big\rangle_{x'},
\end{eqnarray}
which indicates that $\langle u_{x2} \rangle_x \neq 0$ whenever ERK signal distrubition has a point $\partial\Delta E/\partial x' \neq 0$. Finally, by converting the coordinates from the moving frame to the experimental frame ($x' \to x$ ${\rm and} \ \partial/\partial x' \to \partial/\partial x $), we then perform the calculation of the Gaussian shape of $\Delta E$ at $y=0$ on the propagation path of the ERK wave. Migration velocity $\bm{u}_{mig}= (\langle u_{x2} \rangle_x,0)$ at $y=0$ is
\begin{equation}\label{vel2}
\bm{u}_{mig} = - \bm{u}_l \frac{\beta \gamma}{2} (\Delta E_0)^2.
\end{equation} 
This form of $\bm{u}_{mig}$ indicates nonzero net motion under the propagating ERK wave (Fig. \ref{fig1}(a) and (c)). $\beta \gamma > 0$ indicates that the cell population migrates in the opposite direction of the approaching signal wave and persistently drives the net motion in one direction. Note that $\beta$ appears in Eq.  \eqref{vel2} reflects the response of the friction coefficient to the ERK change (Eqs. (13)-(15)), while $\gamma$ in Eq. \eqref{vel2} comes from the response of the ERK-dependent cell density (Eq. (16)). Therefore, these two parameters have different physical origins, and their product determines the speed of the ERK wave-driven collective migration.

\section{ERK signaling in MDCK cell monolayer as an experimental model}

We experimentally verified the theoretical model using monolayer-forming epithelial cells. We used Madin-Darby canine kidney (MDCK) cells because they exhibit collective migration coupled to intracellular signaling activation. In a monolayer of MDCK cells, the activity of ERK MAP signaling (ERK signaling) propagates like a traveling wave and drives collective migration in the opposite direction of the signal wave \cite{MATSUBAYASHI2004, Aoki2017, hino2020}. During collective migration, ERK signaling regulates contractility through the myosin molecular motor. ERK propagates through the cell monolayer as a traveling wave, exhibiting rectified migration. The presence of cells with high ERK activity in a cell population can also induce spatial changes in the contractile force gradient and orientation order between cells. Consequently, we considered an ERK-regulated MDCK cell monolayer to be an appropriate experimental system for our theoretical model.

The MDCK cells used in the experiment were optogenetically controlled through the ERK pathway using light-induced dimerization (LID) of CRY2-CIBN \cite{AOKI2013, Aoki2017} (Fig. \ref{fig2}(a)) with 488 nm blue LED laser. The phosphorylation activity of ERK MAP kinase increased in response to light intensity, during which ERK-mCherry protein was translocated to the nucleus (Fig. \ref{fig2}(b)). This tool allowed for manipulating the ERK activity at the single-cell level in a light-responsive manner. Because the ratio of the amount of ERK-mCherry partitioned between the cytoplasm and nucleus is proportional to ERK activity in this setup (Fig. S1 \cite{supplement}), the level of activity was quantitatively assessed by measuring the nuclear translocation of ERK-mCherry proteins (Fig. \ref{fig2}(c)). The detailed calculation method of ERK activity is provided in the Supplemental text \cite{supplement}. 

We tested whether light could regulate ERK signaling activity and thereby induce changes in MDCK cell density and orientation within the monolayer. To perform this analysis, we included a light-unresponsive cell population to create differences in signal intensity between cells. We made a mosaic cell population by mixing the optogenetic MDCK cells with normal cells at a 1:9 ratio. This heterogeneous cell population was then exposed to blue light, which induced ERK signaling only in light-responsive cells. After applying optogenetic activation for 30 min, we tested whether the activated cells exhibited an increase in the cell area. The ratio of changes in cell area is a non-dimensional parameter defined as $(A_1 – A_0)/A_0$, where $A_0$ and $A_1$ are the cell area before and after optogenetic ERK activation, respectively. 

\add{In Fig. \ref{fig2}(d), ERK activation in our mixed MDCK population produces a dose-dependent increase in the fraction of area occupied by cells. As light intensity (and hence ERK activity) rises, the cell‐covered area grows (Fig. \ref{fig2}(g)). In contrast, the group of wild-type MDCK cells, which do not respond to the light stimulus, shows no such trend under identical illumination (Fig. \ref{fig2}(g)). In addition, a single optogenetic cell isolated from its neighbors maintains a constant cell‐occupied area regardless of ERK activation level (Fig. \ref{fig2}(e)), and isolated wild-type cells likewise fail to change their occupied area in response to light (Fig. \ref{fig2}(h)).} These results suggest that the cell area has expanded due to forces generated by the activated cells pulling on the inactive cell. 

We also found with the aid of traction force microscopy that intracellular contractility decreased in MDCK cells with elevated ERK activity (Fig. S3) \cite{ueki2015, supplement}, indicating that cells with increased ERK activity have decreased contractility. The surrounding cells maintain their contractility, resulting in an enlarged area due to the pulling from surrounding cells. Thus, because the cell spreading area is increased by ERK activity, local activation of ERK signaling results in reduced cell density, $\gamma <0$.

We also investigated whether the isotropy of the cell orientation is influenced by ERK activation by inducing the light stimulation to optogenetic MDCK and analyzing the cell orientation order. The orientation order parameter before light exposure was set to $S_0$; after optogenetically upregulating ERK activity for 30 minutes, the difference in the orientation order $\Delta S=S-S_0$ was analyzed at different levels of ERK activity. As ERK activity increased, the rate of change in the orientation order decreased (Fig. \ref{fig2}(f) and Fig. S4), indicating that the orientation of the cell population became more random due to ERK activation. 

\add{Additionally, to verify that the observed ERK activation via optogenetics is not a direct response of MDCK cells, we performed control experiments on both confluent monolayers and isolated single wild‐type MDCK cells lacking the LID pathway. In these controls, light illumination (laser-diode current 40 mA, Fig. \ref{fig2}(b)) produced no changes in any of the following metrics: cell area occupancy in the population (Fig. \ref{fig2}(g)), area of isolated single cells (Fig. \ref{fig2}(h), or orientational order in the population (Fig. \ref{fig2}(i)). These results confirm that the light‐induced effects reported in Figs. \ref{fig2}(d) and (f) arise specifically from optogenetic control of ERK activity. We note that a mix of optogenetic MDCK and normal MDCK cells was only used for the experiments shown in Fig. \ref{fig2}(d)-(i)}.

Previous studies on MDCK cell populations reported that the velocity tends to increase when cells move parallel to the orientation angle \cite{saw2017, balasubramaniam2021investigating}. In the present study, on the contrary, we introduced the relationship between the changes in orientation order and friction, as seen in Eq. \eqref{friction}. The decrease in friction in the direction of alignment corresponds to a negative friction parameter, $\epsilon <0$ (the opposite is observed in neural progenitor cells, where $\epsilon >0$ \cite{kawaguchi2017}). According to $\beta \sim - \epsilon\frac{dS}{dE}$, the change in the friction coefficient in our experiment ($\epsilon < 0$, $\frac{dS}{dE}<0$) was $\beta <0$. 

\section{Collective migration under a synthetic ERK wave}

\begin{figure*}[tb]
\begin{center}
\includegraphics[width=12cm]{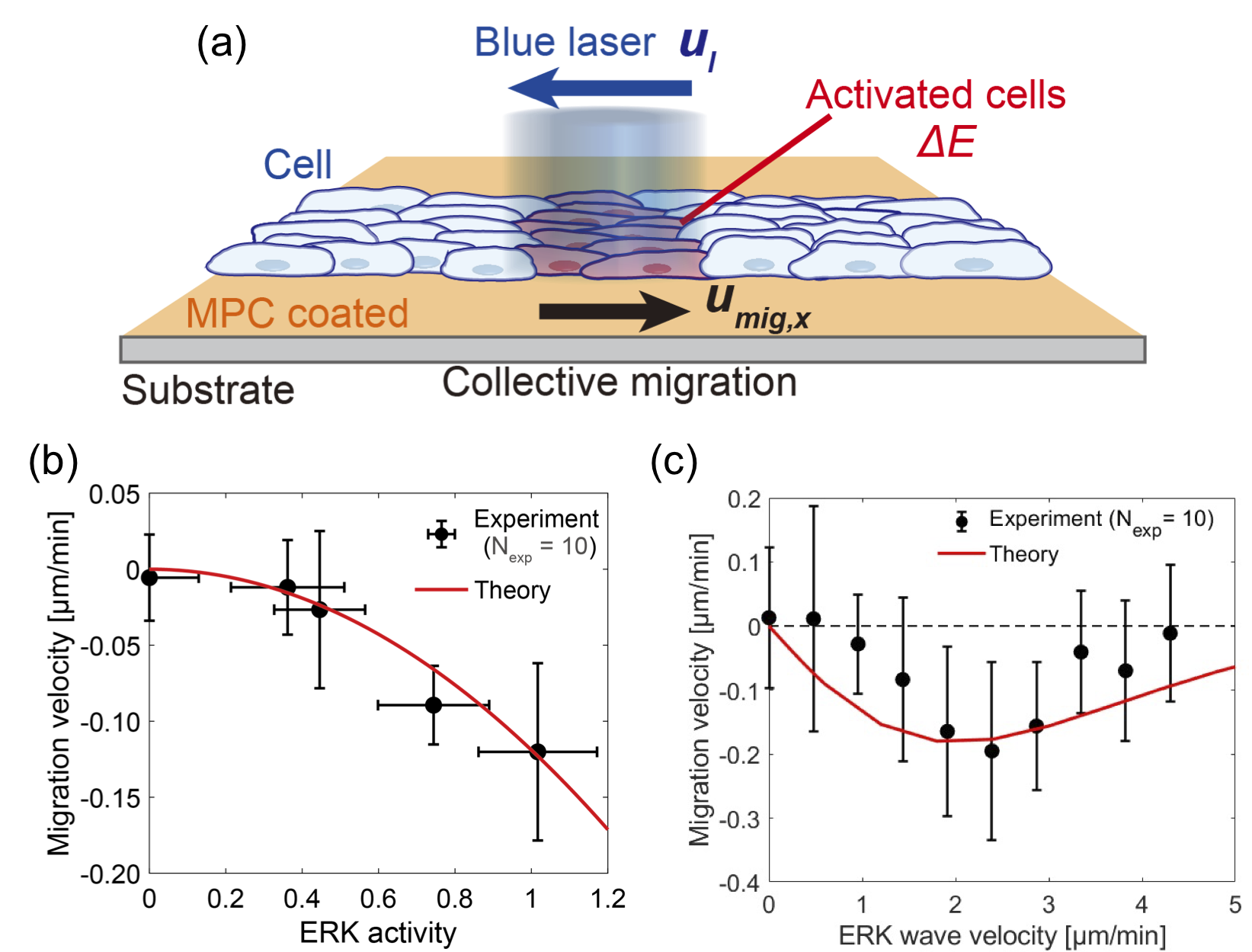}
\caption{\textbf{Directed collective migration of MDCK cells guided by a synthetic signal wave}. (a) Collective migration by a sweeping synthetic ERK wave. The cell monolayer with the LID pathway was patterned into a channel-shape of width \SI{500}{\micro\meter}. A spot of blue light was moved along the cell monolayer at a velocity of $\bm{u}_l$. (b) Squared dependence of the migration velocity on the magnitude of the synthetic ERK wave. \Adddd{The theoretical curve was drawn by fitting with Eq. \eqref{vel2} with $\beta\gamma = 0.14$ (95\% CI is 0.00 to 0.28). The number of technical replicates is $n=2$, and the number of independent experiments is $N_{exp}=10$. The error bars in horizontal and vertical axes indicate the SD.} (c) Wave-speed dependent collective migration. The maximum velocity of the migration occurred at an intermediate ERK wave speed (black circles). \Adddd{The theoretical curve of Eq. \eqref{mig2}, which also includes the calculation according to Eq. \eqref{erk2}, was fitted with $\beta \gamma$ = 0.12 (95\% CI is 0.00 to 0.25) and fixed parameters $\tau_d$ = 40 min, $\lambda$ = \SI{450}{\micro\meter}, and $\tau_E$ = 15 min \cite{AOKI2013}(solid red lines in (b) and (c)). The number of technical replicates is $n=2$, and the number of independent experiments is $N_{exp}=10$. The error bars indicate the SD.}}\label{fig3}
\end{center}
\end{figure*}

Experiments on the optogenetic control of ERK suggest that MDCK cells respond to an increase in ERK signaling with a friction change of $\beta <0$ and density change of $\gamma <0$. Eq. \eqref{vel2} indicates that $\beta \gamma>0$, suggesting that the motion is in the opposite direction to the ERK signal wave. To test this theory, we designed a synthetic ERK wave mimicking a natural signal wave through the unidirectional sweeping of a focused laser light (Fig. \ref{fig3}(a)).
MDCK cell population was confined in \SI{500}{\micro\meter} wide channel-shaped region. This confinement size is only slightly larger than the velocity correlation length of the MDCK cell monolayer (that is approximately \SI{300}{\micro\meter} \cite{shigeta2022}), and is large enough to achieve a one-dimensional system without strong effects of constraints. The area covered by the laser light used for optogenetics is  $\sim$\SI{100}{\micro\meter} in diameter, and spatial differences in ERK activity are produced with and without light stimulation. Therefore, the situation shown in Figure 2(d), where cells with high ERK activity are surrounded by cells with low ERK activity at the basal level, is realized by this optogenetic manipulation.

The synthetic ERK wave travels at a constant speed ${u}_l = $\SI[per-mode=symbol]{2.0}{\micro\meter\per\minute}, whereas $\Delta E$ is regulated by adjusting the laser intensity. The velocity of cell migration was always negative, indicating that the cell population with increased ERK activity moved in the direction opposite to that of the ERK wave (Fig. \ref{fig3}(b)). Moreover, as expressed in Eq. \eqref{vel2}, the collective migration speed $|\bm{u}_{mig}|$ is proportional to the square of ERK activity, $(\Delta E)^2$. In the experiment, the migration velocity exhibited a quadratic increase with $\Delta E$ (Fig. \ref{fig3}(b)). Our theoretical model is thus consistent with experimental observations; in particular, the quadratic increase in migration speed implies that the two ERK-dependent processes of density and friction are involved in rectifying collective migration under the traveling signal wave.

\begin{figure*}[tb]
\begin{center}
\includegraphics[width=17cm]{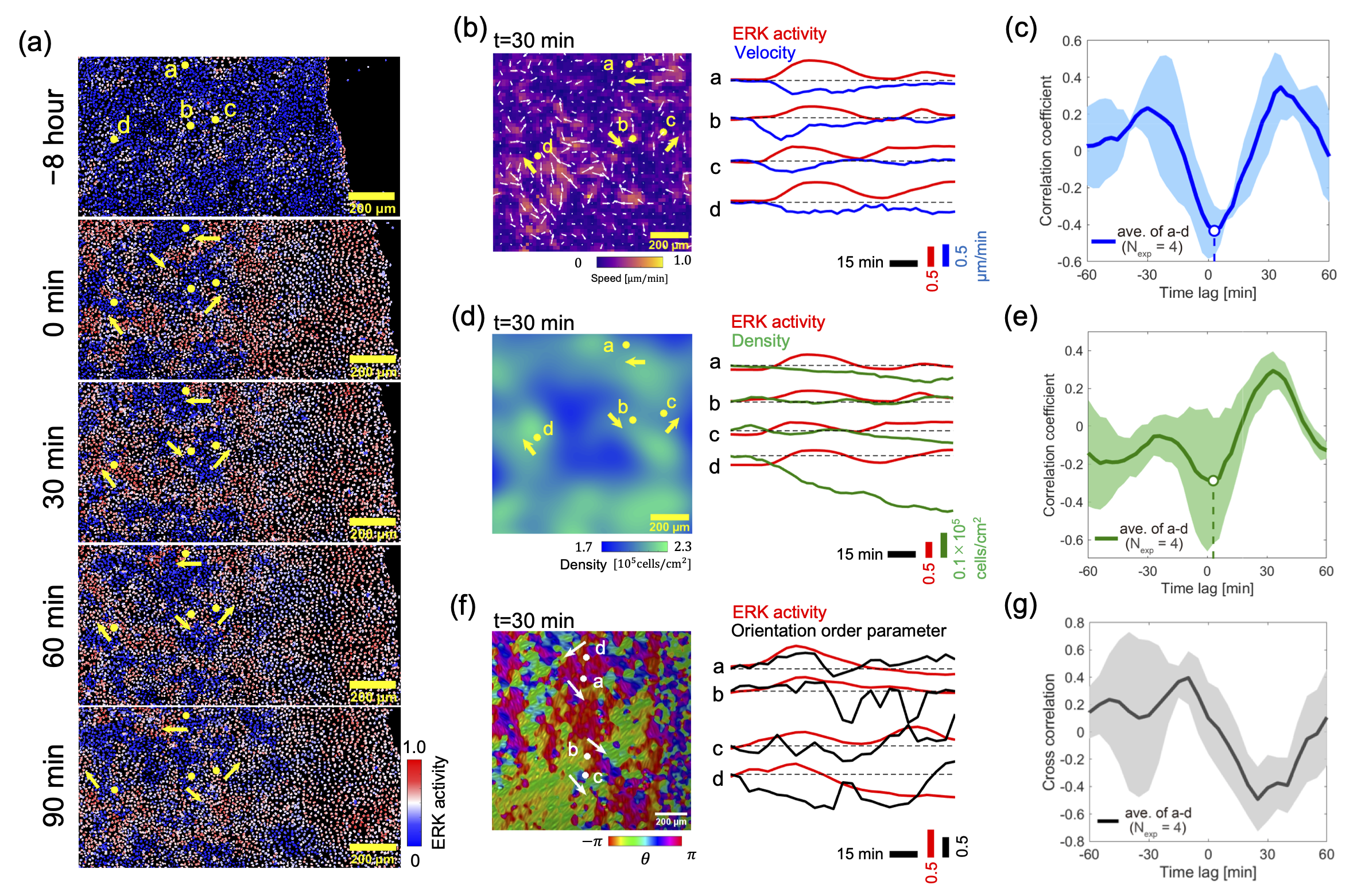}
 \caption{\textbf{ERK wave-driven collective migration in the wound healing process of MDCK cells.} (a) Spatiotemporal alterations in ERK activity within the wound-healing MDCK cell population, as measured using a FRET probe, with red indicating heightened levels of ERK activation. The yellow dots indicate the positions where the time course of ERK activity, migration velocity, orientation order, and cell density are plotted as representative data in Fig. 4(b), (d), and (f), respectively. Alphabetical letters are used to distinguish between these points. The yellow arrows indicate the direction of ERK wave propagation, determined by the ERK activity measured using a FRET probe. The time $t=0$ was defined as \SI{8}{\hour} after the start of the wound healing assay. (b) Time course of cell migration speed and ERK activity within the cell monolayer. The reference value at $t=0$, indicated by the dotted line, was used to evaluate the degree of change in migration velocity (density in (d) and orientation order in (f)) from the basal level. (c) Cross-correlation function between ERK activity and cell migration. A negative simultaneous correlation indicates that the collective migration opposing the ERK wave dominates wound healing propagation. (d) Time course of ERK activity and cell density. (e) Cross-correlation function between ERK activity and fluctuations in cell density. A negative correlation at a time lag of $\Delta t=0$ indicates that an increase in ERK activity is correlated with a decrease in cell number density. (f) Cell orientation dynamics. The orientation angle of individual cells within the cell monolayer is displayed through a color code. Changes in the orientation order parameter $S$ are plotted with ERK activity over time. (g) Cross-correlation function between orientation order and ERK activity. A negative correlation value of $\Delta t=\SI{30}{\minute}$ indicates that an increase in ERK activity is associated with a decrease in orientation order. \Adddd{In Figs. 4(b) - (g), four data points ($N_{exp}=4$) were collected in the measurement of the wound-healing cell population, and the number of technical replicates is $n=1$ for these measurements. The image data used for orientation analysis in Fig. 4(f) is different from those used for Figs. 4(b) and 4(d).}}\label{fig4}
 \end{center}
 \end{figure*}

We measured the dependence on the ERK wave velocity and found that the maximum migration velocity ${u}_{mig} = \SI[per-mode=symbol]{-0.2}{\micro\meter\per\minute}$ was reached at a wave velocity of ${u}_l = \SI[per-mode=symbol]{2.3}{\micro\meter\per\minute}$ (Fig. \ref{fig3}(c)). Such a wave velocity in directed collective migration has also been reported in previous studies \cite{Aoki2017, lozano2019, hino2020}; however, the mechanical mechanisms remain unclear. Accordingly, we considered viscoelastic deformation and restoration in cell migration. The extended deformation coefficient is $\Gamma_{\tau_d}= \gamma \left[1- \exp \left (-\frac{\lambda}{u_l\tau_d}\right)\right]$, where $\tau_d$ is the characteristic time of mechanical restoration ($\tau_d \approx $ \SI{40}{\minute} \cite{iyer2019}) and $\lambda$ is the wavelength of the ERK wave ($\lambda$ = \SI{450}{\micro\meter}) (see detailed calculation in \cite{supplement}). By considering the relaxation time in the density change in $\Gamma_{\tau_d}$ \cite{hamadi2005}, the migration velocity $\bm{u}_{mig}$ can be extended to the general form
\begin{equation}\label{mig2}
\bm{u}_{mig} = -\bm{u}_l\frac{\beta \Gamma_{\tau_d}}{2}(\Delta E_0)^2.
\end{equation}
$\Gamma_{\tau_d}$ affects the migration speed according to the relationship between the relaxation time of the density ($\tau_d$) and the propagation time of the wave ($\frac{\lambda}{u_l}$). For slowly propagating ERK waves ($u_l / \lambda \ll 1/\tau_d$), the migration velocity ${u}_{mig} \propto u_l(1-\exp[-\lambda /(u_l\tau_d)]) \sim u_l$ and thus migration speed linearly increases with ERK wave speed. Hence, the relaxation process can explain the gradual increase in ${u}_{mig}$ at low wave speeds. 

In contrast, for rapidly propagating ERK waves ($u_l / \lambda \gg 1/\tau_d$), the migration velocity plateaus at ${u}_{mig} \propto u_l(1-\exp[-\lambda /(u_l\tau_d)]) \sim \lambda/\tau_d$. This means that the migration velocity does not decrease at large $u_l$, and we need to consider another relaxation process to explain the non-monotonic increase of the migration velocity. To address this point, we also consider the decay of signaling activity with increasing $u_l$. The relaxation dynamics of ERK signaling are given by
\begin{equation}
\frac{d \Delta E}{d t}=-\frac{\Delta E}{\tau_E} + I_s, \label{erk2}
\end{equation}
where $I_s(\bm{x},t) = I_0\exp(-(\bm{x}-\bm{u}_lt)^2/2a^2)$ is the point source of the signal wave and $\tau_E$ is the relaxation time of the activated signal \cite{Aoki2017,supplement}. The duration of signal activation is defined as the time scale $T \approx b/u_l$, with a typical cell size of $b = \SI{20}{\micro\meter}$. If the activation time $T$ is sufficiently longer than the ERK signal relaxation time $\tau_E$ ($T \gg \tau_E$), $\Delta E$ is fully activated by the saturation level, and no change appears in $\bm{u}_{mig}$ for the slowly propagating ERK wave. However, if the activation time is shorter than the relaxation time ($T \ll \tau_E$) under fast-moving waves, $\Delta E$ decreases, in turn damping the associated changes in contractile stress and reduction in friction over time. Because of such a reduction of $\Delta E$ at large $u_l$, the migration velocity $u_{mig}$ decreases at fast propagating ERK wave (Fig. 3(c)).

Therefore, the interplay between density relaxation and signal activation may determine the ERK wave speed of the fastest motion. Our model predicted such wave velocity as $u_l = 2\sim \SI[per-mode=symbol]{3}{\micro\meter\per\minute}$, which is comparable to the experimental value of \SI[per-mode=symbol]{2.3}{\micro\meter\per\minute} (Fig. \ref{fig3}(c)).

\begin{figure*}[tb]
\begin{center}
\includegraphics[width=16cm]{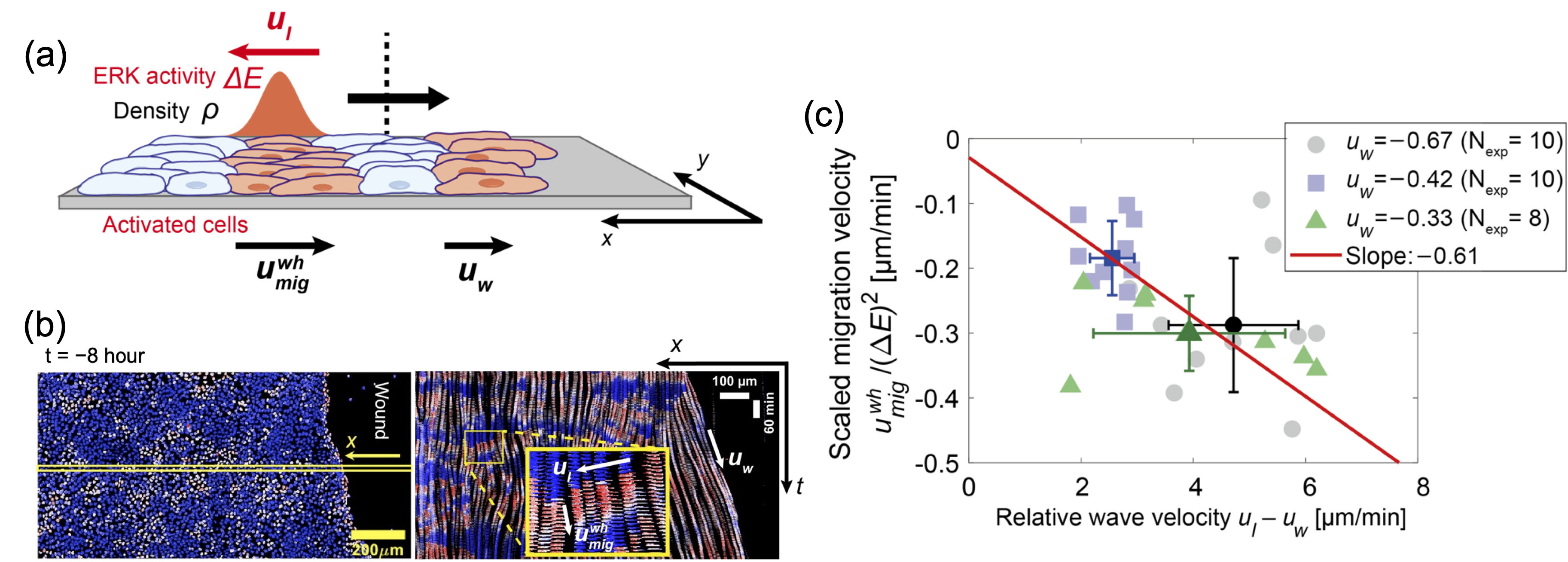}
 \caption{\textbf{Collective migration in a traveling ERK wave during wound healing.} (a) Schematic of collective migration associated with the ERK wave during wound healing. We consider the edge of MDCK cell monolayer move at wound progression $u_w$. \add{$u_{mig}^{wh}$ reflects the local, instantaneous migration of cells in the monolayer interior, whereas $u_w$ measures the global, net advance of the wound boundary.} (b) Quantification of the ERK wave speed $u_l$, migration speed \add{$u_{mig}^{wh}$}, and wound progression speed $u_w$. We measured the speed parameters from the kymograph of MDCK cells undergoing wound healing. In Fig. \ref{fig4}(a), the image at the left shows MDCK cells at $t= - \SI{8}{\hour}$. Cell nuclei were visualized and detected by the fluorescent probe, with the spatial axis toward the wound region on the horizontal axis and the time axis on the vertical axis. The color of the labeled cell nucleus represents the intensity of ERK activation as indicated in the color bar. (c) Relationship between migration speed and ERK wave speed. Data plotted in different colors represent three independent experiments. The ERK activity $\Delta E$ was measured using a FRET probe sensor. ERK activity was normalized against its maximal value. Each data point corresponds to directed migration under one ERK wave. The solid line represents Eq. \eqref{wound} with $\beta \gamma = 0.11$ (95\% CI is 0.02 to 0.23), with fixed parameters $\tau_d$ = 40 min and $\lambda = \SI{450}{\micro\meter}$. The number of technical replicates is $n=3$, and the number of independent experiments is $N_{exp}=8$ or 10.}\label{fig5} 
 \end{center}
\end{figure*}

\section{Wound healing of epithelial cell monolayer}

The proposed model of wave-directed collective migration inspires an experimental investigation of the interplay between naturally occurring signaling waves and the mechanics of the epithelial cell monolayer during wound healing (Fig. \ref{fig4}(a) and \Add{Supplemental Movie 1}). In wound healing experiments, MDCK cell populations were cultured in internal compartments of removable chambers to obtain monolayers, and the compartment walls were removed when they reached a confluency that indicated collective migration from the leading edge \cite{Aoki2017}. This allowed the \textit{in vitro} wounded areas to spread to those where cells were not present, defined as an injured region. We obtained time-lapse images for quantitatively measuring ERK signaling. The level of ERK activation was reported with a FRET sensor of ERK phosphorylation, which was expressed in the cell nucleus, and both cell motility and ERK activity levels were simultaneously measured by time-lapse fluorescence microscopy (see supplemental materials \cite{supplement}). ERK activity remained high in the leading edge of the cell population during wound healing, and ERK wave generation occurred far away. Therefore, the analysis of ERK waves was targeted to the region \SI{500}{\micro\meter} away from the wound edge. Upon \Addd{the removal of the compartment walls from the cell monolayer}, directed collective cell migration is initiated, leading to wound closure. The initiation of collective migration applies mechanical stimuli to the leading cell layer, which activates ERK signaling. The FRET sensor for ERK activation showed that the highly active region (red cells) was localized near the wound site, indicating that ERK signaling propagated from the leading edge of the wounded cell population (Fig. \ref{fig4}(a)). It has been shown that cells collectively migrate in the direction opposite to the propagation of the ERK wave and then fill the existing wound starting from the leading edge \cite{Aoki2017, hino2020, Hiratsuka2015}. By measuring the intensity of ERK signaling and cell migration relative to the direction of ERK wave propagation, we observed that the speed of cell motility increased as the intensity of the signal decreased (Fig. \ref{fig4}(b)), indicating a negative correlation between motility speed and changes in ERK signaling (Fig. \ref{fig4}(c)). We examined changes in cell density associated with ERK activity during wound healing. \Add{Cell density was evaluated based on the number density of cells in a square with \SI{101}{\micro\meter} side length. Although the theoretical model considers the cell's mass density, it is represented using the ratio of the area occupied by cells in the experiment. We found that local cell density decreased with increased signaling (Fig. \ref{fig4}(d) and (e)), consistent with previous findings \cite{Aoki2017, saraswathibhatla2022}.}

We analyzed the orientational dynamics of MDCK cells induced by ERK signaling. The orientation angle $\theta(\bm{x})$ in the cell monolayer was calculated from bright-field images (Fig. \ref{fig4}(f), left)\cite{supplement}. \Adddd{The orientation field was then used to calculate the scalar order parameter $S(\bm{x})= \sqrt{\langle \cos2 \theta \rangle_{ROI}^2 + \langle \sin2\theta \rangle_{ROI}^2 }$ by averaging over the region of interest (ROI).} We analyzed the signal-dependent change in $S(\bm{x})$ and found that this order parameter decreased as ERK activity increased (Fig. \ref{fig4}(f), right). We also calculated the correlation between ERK signaling and $S(\bm{x})$ and observed a negative cross-correlation function with a time delay $\Delta t=\SI{30}{\minute}$ (Fig. \ref{fig4}(g)), indicating that local orientation becomes less organized as ERK activity increases. 
For a typical ERK wave size of $\lambda =$ \SI{450}{\micro\meter} and an ERK wave velocity of $u_l \approx \SI{2.0}{\micro\meter\per\minute}$, cells are exposed to ERK activity for an activation time $T \approx 220$ min activation. From the time delay $\Delta t \approx \SI{30}{\minute}$ shown in Fig. 4(g), this duration of ERK activation is significantly longer than the relaxation time of the $Q$ tensor ($\approx 30$ min).  This indicates that the spatial distribution of the $Q$ tensor becomes stationary upon activation by the ERK wave. Assuming that the local orientation order is related to anisotropic friction, the ERK signal wave may facilitate the reorientation of cells along the direction of wave propagation, making it easier for the cell population to move in that direction.

To extend our theoretical model to wound healing \cite{Ilan2014, basan2013, brugues2014, tetley2019}, we considered a cell monolayer whose leading edge can move in one direction (Fig. \ref{fig5}(a)). We assumed that the ERK signal propagates from the wound site at speed $\bm{u}_l$ within the cell monolayer. \add{Two distinct measures of collective migration in wound healing are defined. First, $\bm{u}_{w}=(u_w, 0)$ is the net wound-closure speed, defined as the time derivative of the average wound-edge position. This motion of the leading cells induces an outward density flux in the cells, $\bm{J}_w = \rho \bm{u}_w$ (Fig. \ref{fig5}(b)). Second, $\bm{u}_{mig}^{wh}$ is the local migration speed of cells in the bulk monolayer as the leading cells move toward the wound region at the constant speed. The cell population then moves at $\bm{u}_{mig}^{wh}$ against the protein signal wave generated from the wound.} Given that the cell population maintains a continuous cell monolayer, Eq. \eqref{continuum} is rewritten as $\frac{\partial \rho}{\partial t} + \bm{\nabla}\cdot(\rho \bm{u}_{mig}^{wh}) = - \bm{\nabla}\cdot \bm{J}_w$. In the moving frame of a traveling signal wave, the density flux is \Adddd{$-(u_l-u_w) \frac{\partial \rho}{\partial x'} + \frac{\partial }{\partial x'}(\rho u_x)+\frac{\partial }{\partial y}(\rho u_y)=0$,} showing that the effective velocity of the signal wave is $\bm{u}_l - \bm{u}_w $ in a system at the tip of the signal wave. The migration velocities in Eq. \eqref{mig2} can be given by \Adddd{(see detailed calculation in \cite{supplement})}
\begin{equation}\label{wound}
\bm{u}^{wh}_{mig} = - (\bm{u}_l - \bm{u}_w)\frac{\beta \Gamma_{\tau_d}}{2}(\Delta E_0)^2.
\end{equation}
Eq. \eqref{wound} indicates that the migration speed is a multiplicative expression as a product of $\beta$ and $\Gamma_{\tau_d}$, with the only difference being that the wave velocity is $\bm{u}_l - \bm{u}_w$. We compared the theoretical models with wound healing experiments by measuring ERK wave velocity and analyzed the rate of leading-edge spreading. We plotted the migration velocity scaled by the ERK signal level $\bm{u}^{wh}_{mig}/(\Delta E_0)^2$ for the stretching velocity of the ERK wave $\bm{u}_l - \bm{u}_w$ (Fig. \ref{fig5}(c)), and found a linear relationship as proposed by Eq. \eqref{wound}. The slope of the curve was expressed as the product of the coefficients of the ERK-dependent density change $\beta$ and the friction change $\gamma$. Here, $\beta\gamma = 0.11$ was comparable between the wound healing and optogenetic manipulation experiments ($\beta\gamma = 0.12$ in Fig. \ref{fig3}(c)).

\section{Discussion}
This study showed that the coordination of changes in cell density and friction as a function of signal activation-induced net collective migration was rectified against a traveling signal wave. \add{ERK activation under the signal wave actually weakens cellular contractility (Fig. S3), leading to a local decrease in cell density directly beneath the wave (Fig. \ref{fig4}(e)).  The signal-dependent active stress causes a local density change, and density flow is generated by the tug-of-war between neighboring cells. In addition, at the front end of the wave, cell stretching accentuated the velocity of motion in the direction opposite to the wave. These ERK-dependent effects collectively drive directed collective migration, whose speed scales quadratically with ERK signal amplitude, ${u}_{\mathrm{mig}} \propto (\Delta E)^2$.}

\add{To describe this behavior, we developed a continuum model of collective migration, rooted in two fundamental relations: momentum conservation of Eq. \eqref{stokeseq} and the continuity equation of Eq. \eqref{continuum}. In constructing this model, we approximated the ERK wave as an isotropically diffusing signal across the monolayer, while other intracellular signaling components were assumed to equilibrate instantaneously. Two experimentally confirmed mechanical responses to ERK activation (Fig. \ref{fig2}(d)) form the foundation of the model: (i) the decrease of local cell density (the increase in individual cell area), captured by the dimensionless density-change parameter $\gamma$ in Eq. (16), and (ii) a loss of orientational order in the monolayer (Fig. \ref{fig2}(f), Fig. S4), motivating the introduction of orientation-dependent anisotropy in the intercellular friction coefficient}. 

\add{We assumed that signal-dependent mobility is governed by the directional dependence, rather than the magnitude, of the friction coefficient in Eq. (4). This assumption is supported by our measurements of adhesion strength in single cells with elevated ERK activity, which showed no significant changes in adhesion magnitude (Fig. S5; see also \cite{tanaka1,tanaka4}). This suggests that cell–substrate friction is not directly modulated by ERK signaling. The remaining plausible mechanism is that the alignment of cells influences how easily they migrate within a monolayer. However, experimentally detecting such frictional anisotropy remains technically infeasible: direct measurement of the intercellular resistance encountered by a cell dragged by external forces, particularly including active force contributions, exceeds the capabilities of current techniques. Despite this limitation, our continuum model offers a mechanistic hypothesis that frictional anisotropy is a central driver of ERK-wave–guided collective migration.}

Theoretical models proposed in previous studies \cite{boocock2021,hannezo2023} have elucidated how cell monolayers can establish polarity in cell migration, attributed to the temporal delay in stress response relative to changes in ERK activity. As the feedback mechanism between ERK wave signaling and cellular stress fields, the self-organized ERK waves can be stabilized during wound healing. Moreover, another theoretical model \cite{Aoki2017, asakura2021hierarchical} suggested that ERK activation affects the frictional force of cell motility by assuming a change in magnitude. In contrast, our study introduces the concept of nematic order in cell alignment and orientation-dependent friction coefficients, replacing wave-induced polarization. Building on recent work linking contractile stress to polarized cell orientation \cite{hino2020,jain2020}, deciphering the molecular pathways by which ERK signaling reorients cells remains an important goal for future study \cite{gardel2022}.

We also found that the existence of the ERK wave speed at which the migration speed is maximal is explained by the interplay of two relaxation dynamics of the cell density change and the decay of the ERK signal. Although we assume that the rate of decay of signaling is faster than that of density change, it is an important question to experimentally test whether the two relaxation times are different enough to explain the ERK-wave speed dependence. A further understanding of the mechanical coordination of contraction and friction, including force-signaling feedback \cite{boocock2021}, may provide physical insights into the active dynamics organized by a traveling signal wave. 

\add{In addition, the continuum model in this study does not consider cell rearrangement within the monolayer.} Therefore, we formulated the relaxation dynamics using viscoelastic relaxation based on the Voigt model \cite{supplement}. Previous research has reported that when epithelial cells rearrange to overcome column obstacles, they behave like an effective viscous fluid following the Maxwell model \cite{tlili2020migrating}. It should be noted, however, that in our wound-healing assay, migrating cells do not experience confinement by obstacles, but rather recover their original cell orientation after complete relaxation from ERK signaling activation, resulting in an elastic response.

\add{Optogenetics offers precise temporal and spatial control of ERK signaling, it does so by replacing the endogenous pathway with an artificially controllable signal. This technical feature, while advantageous for mechanistic studies and direct testing of theoretical predictions, limits the applicability of optogenetic manipulation for probing spontaneous or naturally occurring collective migration. Accordingly, in the latter part of this study, particularly in the wound healing experiments, we did not employ optogenetic control. Instead, optogenetic methods were restricted to the controlled generation of synthetic ERK signal waves, allowing us to isolate and examine the effects predicted by our coarse-grained theoretical model under $\textit{in vitro}$ experimental conditions.} 

Understanding how signal transduction systems regulate directed collective migration and ordered pattern formation, not only in epithelial cells but also in other cell types will aid in understanding the principles by which cell populations build functional forms while adapting to complex geometric constraints \cite{deforet2014emergence, shigeta2022, ienaga2023}. Such interplay of mechanical force and protein signaling is involved in apoptosis \cite{gagliardi2021} and tissue regeneration \cite{de2021}. \add{While this study focuses on epithelial cell monolayers and ERK signaling, the underlying biophysical framework would be applicable to a wide range of wave-guided dynamics in other biological contexts. Propagating signal waves have been observed in diverse systems, such as Hes7 transcription factor of segmentation clocks \cite{yaman2023controlling}. The theoretical description developed here may offer relevant insights for understanding and predicting collective migration in various biological systems where wave-like signaling and mechanical coordination are essential.} Therefore, elucidating the physical mechanisms underlying these mechanical-chemical interactions is an important challenge for future research.

\section*{Materials and Methods}

\subsection*{Cell culture and microscopy}
An epifluorescence microscope (Olympus, IX73) equipped with a CMOS camera (Andor, Zyla) was used for microscopic observation. The nuclei were labeled with H2B-iRFP fluorescent protein to track individual cells, and live fluorescent imaging was taken at 5-minute intervals for \SI{8}{\hour}. The cell migration velocity was calculated by dividing the change in the position of the fluorescent protein-labeled cell nucleus by the time interval of \SI{5}{\minute}. For the experiment in Fig. \ref{fig3}, to ensure that cell migration occurs along optogenetically induced ERK signal wave, MDCK cells were seeded in a quasi-one-dimensional channel pattern with the width of \SI{500}{\micro\meter} by a micropatterning method previously reported \cite{ienaga2023}. 

\Adddd{MDCK cell line was cultured in MEM (ThermoFisher Scientific, 11095-072) supplemented with 10\% fetal bovine serum (FBS; Sigma-Aldrich, 172012-500mL), 1x GlutaMax (ThermoFisher Scientific, 35050061), and 1 mM sodium pyruvate (Sigma-Aldrich, S8636-100mL) in a 5.0\% CO$_2$ humidified incubator at \SI{37}{\celsius}. We also used optogenetic MDCK cells (MDCK/CRY2-CRaf/CIBN-EGFP-KRasCT/H2B-iRFP/
FLAG-MEK1-mCFP/mCherry-ERK2KD) established in Ref.\cite{Aoki2017}(Fig. 2(a)). MDCK cells expressing CRY2-CRaf, CIBN-EGFR-KRasCT, H2B-iRFP, FLAG-MEK1-mCFP, and mCherry-ERK2KD were selected with antibiotics and then maintained in a minimum essential medium (MEM, ThermoFisher Scientific). For cultivation of optogenetic MDCK cells, selective antibiotics (\SI{1.0}{\micro\gram\per\milli\liter} puromycin, \SI{80}{\micro\gram\per\milli\liter} G418, \SI{10}{\micro\gram\per\milli\liter} zeocin, and \SI{1.0}{\micro\gram\per\milli\liter} blasticidin S.) were added in the MEM. For time-lapse imaging, the MDCK cells were placed on a glass base dish ($\phi$=\SI{35}{\milli\meter}, IWAKI, 3000-035) with a channel-shaped pattern \SI{500}{\micro\meter} in width with MPC polymer (Lipidure-CM5206, NOF Corporation) on its surface. Figures 4 and 5 show the wound healing assay. We cultured the cell population in a two-well culture insert (ibidi, 81176) and then removed the culture insert when the density reached the confluent level. Microscopic observation was started at the time of removal, and time $t=0$ was defined as 8 hours after removal of the culture insert. The propagation of ERK waves was observed at a distance of at least \SI{500}{\micro\meter} from the area where the wound was formed.}

\Adddd{\subsection*{Quantitative analysis}
We performed fluorescent labeling of the cell nucleus at each time point in the time-lapse measurement, and the velocity of cell migration was determined by calculating the distance traveled per unit time of the cell nucleus. The orientation angle in the cell monolayer at position $\bm{x}$, defined as $\theta (\bm{x})$, was calculated from brightfield images using OrientationJ in the Fiji plugin, with the window size set to \SI{0.75}{\micro\meter} and the gradient type set to cubic spline. The orientation angle was then analyzed by calculating the orientation order parameter of a two-dimensional system, $S=\sqrt{\langle \cos2\theta\rangle_{ROI}^2+\langle \sin2\theta\rangle_{ROI}^2}$ within a region of interest (ROI) of size $w=\SI{50}{\micro\meter}$. $S$ is a scalar order parameter used to analyze the signal-dependent change of the orientation field.}

\Adddd{We evaluated the signaling activity of ERK MAP kinase by recording the translocation of ERK-mCherry into the nucleus. We recorded the fluorescence intensities of ERK-mCherry inside and outside the nucleus, $C_{in}$ and $C_{out}$, respectively. We then defined the ERK signal $E$ as
\begin{equation}\label{erkact}
E=\frac{C_{in}}{C_{out}}.
\end{equation}
The activated ERK signal was calculated as $\Delta E = E - E_0$, where $E_0$ is the ERK signal at basal conditions (the absence of blue laser illumination).}

We analyzed the dynamics of migration speed, density change and orientation change at the location of interest (labeled with \textbf{a}, \textbf{b}, \textbf{c}, \textbf{d}) as the ERK wave propagates. The cross-correlation functions between ERK activity and migration velocity ($C_{u}(\Delta t)$), ERK activity and local cell density ($C_{\rho}(\Delta t)$), and ERK activity and orientation order ($C_{S}(\Delta t)$), which appeared in Fig. \ref{fig4}, are defined as follows
\begin{eqnarray}
C_{u}(\Delta t)&=&\frac{\langle (\Delta E(t) - \overline{\Delta E})(u_x(t+\Delta t) - \overline{u}_x)\rangle_t}{\langle (\Delta E(t) - \overline{\Delta E})(u_x(t) - \overline{u}_x)\rangle_t} \nonumber\\
C_{\rho}(\Delta t)&=&\frac{\langle (\Delta E(t) - \overline{\Delta E})(\rho(t+\Delta t) - \overline{\rho})\rangle_t}{\langle (\Delta E(t) - \overline{\Delta E})(\rho(t) - \overline{\rho})\rangle_t} \nonumber\\
C_{S}(\Delta t)&=&\frac{\langle (\Delta E(t) - \overline{\Delta E})(S(t+\Delta t) - \overline{S})\rangle_t}{\langle (\Delta E(t) - \overline{\Delta E})(S(t) - \overline{S})\rangle_t} \nonumber
\end{eqnarray} 
where $\overline{\Delta E}$, $\overline{u}_x$, $\overline{\rho}$, $\overline{S}$ are the average of ERK activity, migration speed against the direction of the ERK wave propagation, local cell density, and orientation order parameter, respectively. $\langle \cdot \rangle$ is the ensemble average over time. In Figure 4, the cross-correlation function is calculated at four locations, and the average is plotted as a solid line, with the standard deviation shown as a light-colored area.

\section*{Data availability statement}

All data supporting the findings of this study are available within the article and its ESI files.

\section*{Conflict of interests}

There are no conflicts to declare.

\section*{Acknowledgements}
This work was supported by Grant-in-Aid for Scientific Research on Innovative Areas (JP16H00805, JP17H05234, and 18H05427 to YTM, JP19H05798 to KA), Grant-in-Aid for Transformative Research Areas (A) (JP23H04711 and JP23H04599 to YTM), Grant-in-Aid for Scientific Research (B) (JP20H01872 and JP23H01144 to YTM, JP18H02444 and JP22H02625 to KA), Grant-in-Aid for Young Scientists (JP22K14014 to TF), JST FOREST Grant (JPMJFR2239 to YTM), CREST JST Grant (JPMJCR1654 to KA), AMED-CREST Grant (JP20GM0810002 to SK), NIBB Collaborative Research Program (18-355 to YTM), Research Grant from Nakatani Foundation (to YTM), and Joint Research of ExCELLS (23EXC205 and 24EXC206 to YTM).

\clearpage

\setcounter{equation}{0}
\setcounter{figure}{0}
\setcounter{table}{0}
\setcounter{section}{0}
\makeatletter
\renewcommand{\theequation}{S\arabic{equation}}
\renewcommand{\thefigure}{S\arabic{figure}}
\renewcommand{\bibnumfmt}[1]{[#1]}
\renewcommand{\citenumfont}[1]{#1}

\widetext
\section{Experimental methods}

\subsection{Biochemical assay for ERK signal activation}

\begin{figure}
 \begin{center}
  \includegraphics[width=80mm]{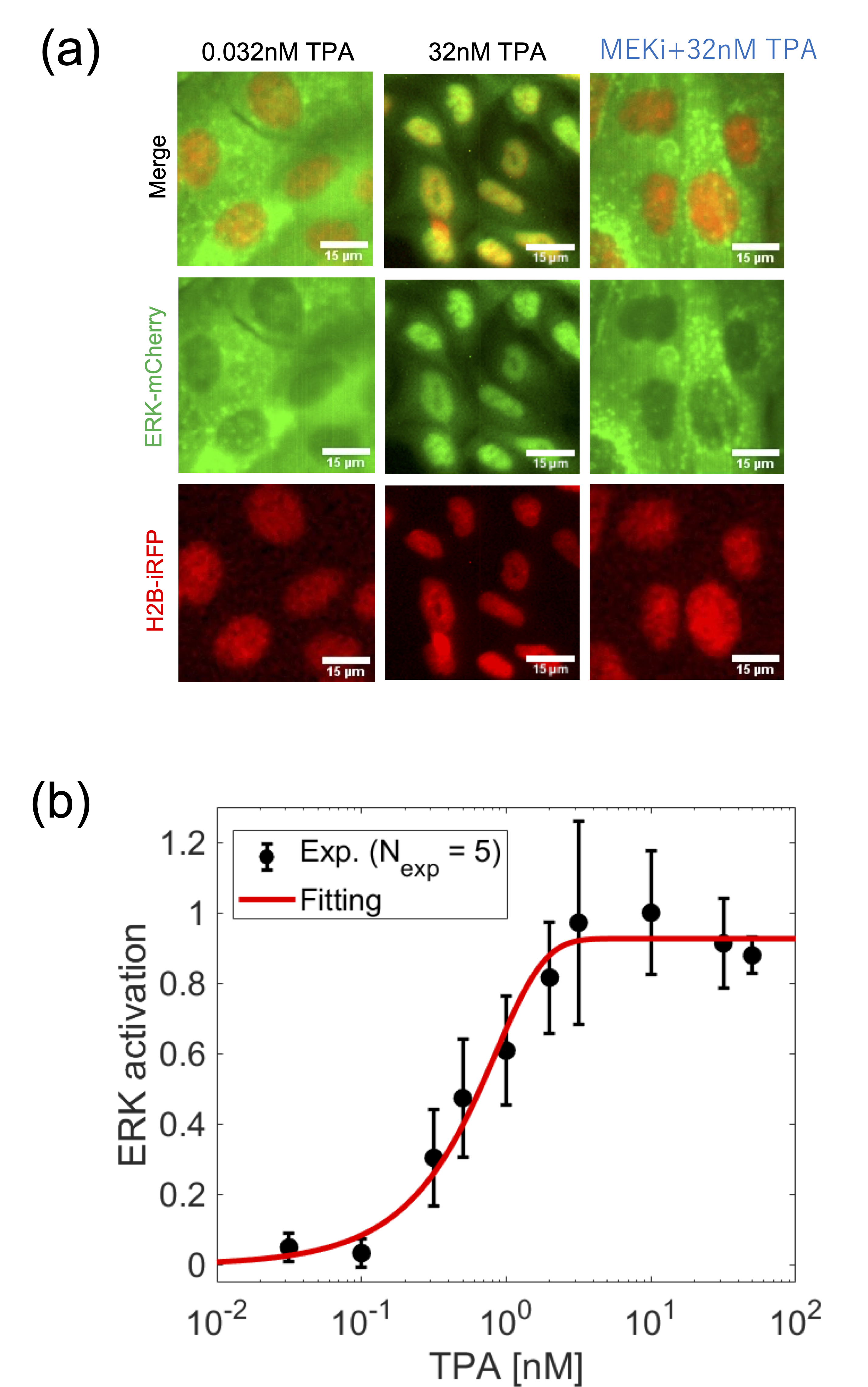}
 \caption{\textbf{Control of ERK activity by TPA chemical treatment.} (a) Quantitative analysis by fluorescence microscopy of the ERK activation by TPA addition. Left column: The addition of 0.032 nM TPA is not sufficient for ERK activation, and ERK-mCherry is abundant in the cytoplasm. Middle column: treatment with 32 nM TPA transduces the signal of the ERK activity, and ERK-mCherry is translocated into the nucleus (marked by H2B-iRFP in red). Right column: the addition of MEK inhibitor suppresses the TPA-induced ERK activation. (b) Calibration curve to show the change in ERK activity depending on the TPA concentration. The black dots are the experimental data, and the red line is the curve fit with the sigmoid function $1/(1+e^{-K_dC})$, with a dissociation constant $K_d=1.06$ and TPA concentration $C$. \Add{Error bars indicate standard deviation (SD). The number of technical replicates is $n=3$ and the number of independent experiments is $N_{exp}=5$.}}\label{fig_TPA} \end{center}
\end{figure}

\begin{figure}
 \begin{center}
  \includegraphics[width=70mm]{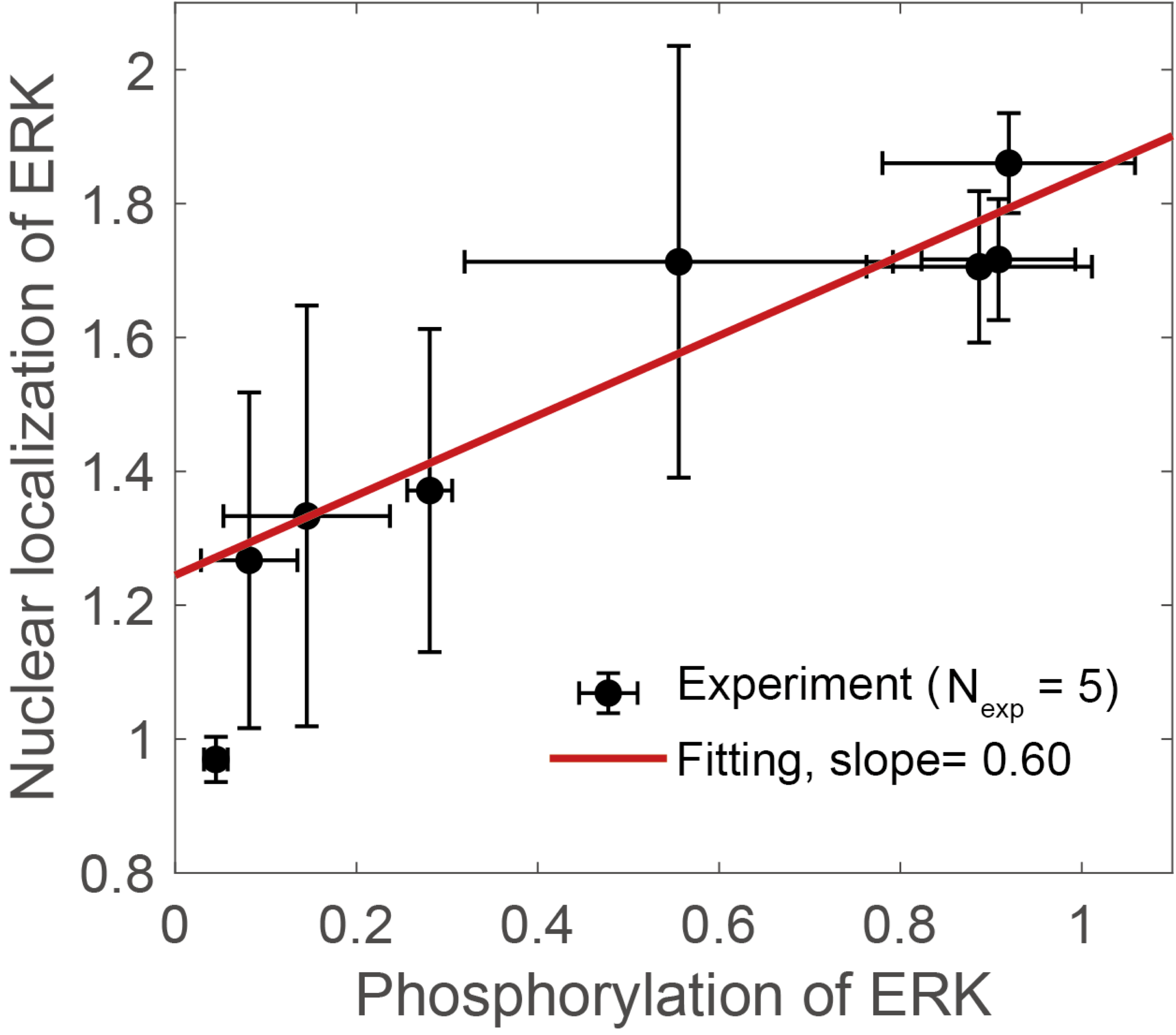}
 \caption{\Add{\textbf{Phosphorylation rate and nuclear translocation rate of ERK MAP kinase protein upon the ERK activation.}  We plot the ratio of fluorescence intensity of ERK in the nucleus and cytoplasm (vertical axis) as a function of the ratio of phosphorylated/non-phosphorylated ERK by Western blotting (horizontal axis). Five points were measured at each TPA concentration (0.032 nM, 0.1 nM, 0.32 nM, 1 nM, 3.2 nM, 10 nM, 32 nM). Black circles indicate the mean values of ERK nuclear translocation and phosphorylation rates at these concentrations, and error bars indicate SD. The red solid line shows the linear fitting, and the slope is $g=0.60$ (95\% confidential interval (CI) is 0.39 to 0.82). The number of technical replicates is $n=3$, and the number of independent experiments is $N_{exp}=5$.}}\label{fig_erkcalb} \end{center}
\end{figure}

In addition to the optogenetic control, we controlled ERK activity with the chemical inducer 12-O-Tetradecanoylphorbol 13-acetate (TPA). In TPA-treated cells, ERK-mCherry in the cytoplasm migrates into the nucleus (Fig. \ref{fig_TPA}(a)), and the rate of nuclear translocation reports the ERK signal activity. We also measured the basal level of ERK signal $E_0$ from the control experiment (0 nM TPA). According to Eq. (29) in main text,  the activated ERK signal was obtained by $\Delta E = E - E_0$. Fig. \ref{fig_TPA}(b) shows a quantitative measure of ERK activity. We determined the positive correlation of the nucleus ERK-mCherry concentration with the phosphorylation activity of ERK at various TPA concentrations. In addition, the ERK-phosphorylated substrate was quantitatively analyzed by Western blot. We confirmed that the level of nuclear translocation was proportional to the level of phosphorylation (Fig. \ref{fig_erkcalb}). 

Moreover, to test whether the changes induced by TPA treatment were specific to ERK signaling, we also performed additional experiments with MEK inhibitor PD0325901, which inhibits ERK activation upstream of the signal pathway. Simultaneous inhibition measures only the ERK-independent effects induced by TPA treatment. 

\subsection{Traction force microscopy}

To measure changes in cell force due to increased ERK activity, we performed traction force microscopy to verify the increase or decrease in force according to ERK activity level.

\begin{figure*}[tb]
 \begin{center}
  \includegraphics[width=180mm]{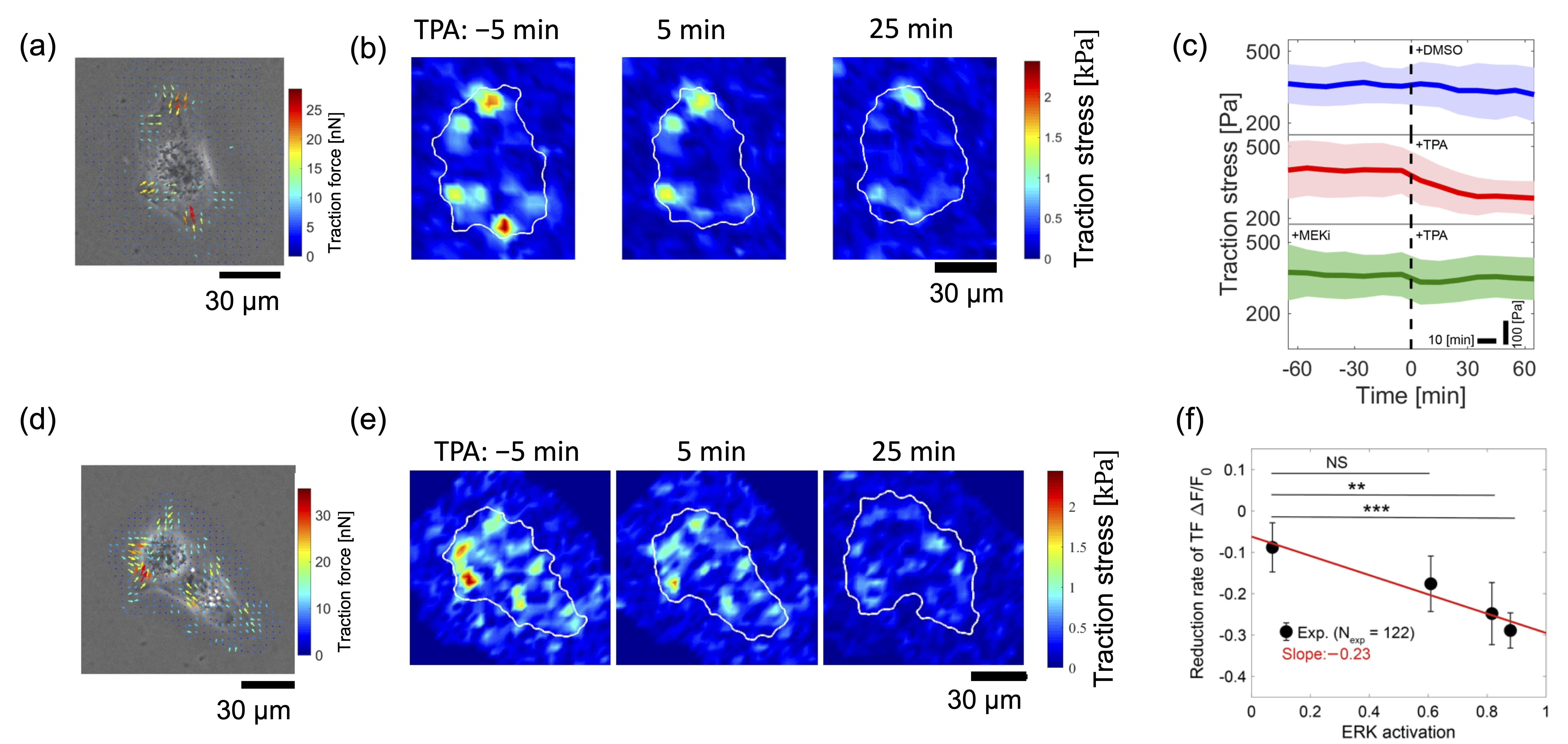}
 \caption{\textbf{The signal activation reduces the traction force of MDCK cells.}  (a) A representative force map is shown with a typical phase contrast image of the single MDCK cell. The scale bar is \SI{30}{\micro\meter}. (b) Time-evolution of the traction stress in a single MDCK cell. Chemical activation by \SI{32}{\nano M} TPA was performed at time $t$ = \SI{0}{\minute}. The absolute value of the traction force is shown as a pseudo-color map. The number of these spots and the traction strength in these spots decreases with the ERK activation. (c) Time evolution of traction force in single MDCK cells. Chemical activation by \SI{32}{\nano M} TPA was performed at time $t = 0$ min. Top: control group (blue, DMSO, $N_{exp}=25$), middle: ERK activation group (red, TPA, $N_{exp}=32$), Bottom: ERK suppression group (green, TPA and MEK inhibitor, $N_{exp}=39$). The solid line represents the average value, and the region in pale color represents the standard deviation. The number of technical replicates is $n=1$ for each case. (d) A representative force map is shown with a typical phase contrast image of the group of adhered cells. The scale bar is \SI{30}{\micro\meter}. (e) Time evolution of the traction stress of the group of cells after ERK activation by TPA treatment (\SI{32}{\nano M} TPA). Like the decrease in traction in a single cell, the number of spots where traction stress accumulates at a higher level decreases with ERK activation. (f) The traction force at various levels of signal activation with \SI{1.0}{\nano M} (56\% ERK activation),  \SI{3.2}{\nano M} (81\% ERK activation), and \SI{32}{\nano M} (88\% ERK activation). Statistical analysis was done by U-test; ** denotes $p < 0.01$ and *** denotes $p < 0.001$ in significance. \Add{The red line shows a linear fitting curve, with a slope of $g=-0.23$ (95\% CI of $-0.34$ to $-0.10$). Error bars indicate SD.}}\label{figs3}
 \end{center}
 \end{figure*}

Photocurable styrenated gelatin (StG) was used as a substrate \cite{ebata2020,ueki2015}. StG (30 wt\%) and sulfonyl camphorquinone (2.5 wt\% of gelatin; Toronto Research Chemicals, ON, Canada) were dissolved in phosphate-buffered saline (PBS). The mixed solution was centrifuged (MX-301; TOMY, Tokyo, Japan) at 14,000 rpm (17,800 g) at \SI{30}{\celsius} for \SI{1}{\hour}, and the deposit was removed. The clear sol solution was aspirated for \SI{20}{\minute} at room temperature to exclude dissolved oxygen. Then, the sol solution was conditioned for 10 min using an AR-100 deforming agitator. 

To embed the fluorescent beads near the surface of the gels, a glass substrate was coated with poly(N-isopropylacrylamide) (PNIPAAm, Sigma Aldrich, St. Louis, MO), StG sol, and StG sol with fluorescent beads (Fluorospheres  Carboxylate-Modified Microspheres, \SI{0.2}{\micro\meter}, red fluorescent (580/605), Invitrogen) by using a spin coater \cite{ebata2021avoiding}. We added 0.1 wt\% TWEEN20 to coat the StG sol solution to improve wettability. 25 $\SI{}{\micro L}$ of the StG sol solution was spread between the vinyl-glass and the coated glass substrate. The gelation of the StG sol was then induced by irradiation with visible light for 360 -- 450 s (45 -- \SI{50}{\milli\watt\per\square\centi\meter} at 488 nm; light source: MME-250; Moritex Saitama, Japan). The hardened gel was detached from the PNIPAAm-coated glass substrate and washed thoroughly with PBS at \SI{28}{\celsius}. 

The elasticity of the StG gel was measured by nano-indentation analysis using atomic force microscopy (JPK NanoWizard 4, JPK Instruments). A commercial silicon-nitride cantilever with a nominal spring constant of 0.03-\SI{0.09}{\newton\per\meter} was used (qp-BioAC-CI CB3, Nanosensors). The elasticity of the gels was approximately 80-\SI{120}{\kilo\pascal}. 

To improve the adhesion between the cells and the gel, the StG gel was chemically modified by fibronectin through N-Ethyl-N0-(3-dimethylaminopropy)carbodiimidehydrochloride (EDC) and hydroxy-2,5-dioxopyrolidine-3-sulfonic acidsodium (NHS). The EDC and NHS were dissolved in 2-(N-morpholino)ethanesulfonic acid (MES) buffer. \SI{750}{\micro\liter} of \SI{100}{\milli M} NHS solution and  \SI{400}{\milli M} EDC solution were mixed. The StG gel was activated by soaking in the NHS/EDC solution for \SI{1}{\hour}  at room temperature. After removing the NHS/EDC solution, the StG gel was washed with PBS two times. Then \SI{0.5}{\milli\gram\per\milli\liter} of fibronectin was dissolved in PBS on the StG gel. The StG gel was then kept at  \SI{4}{\celsius} for \SI{24}{\hour}. After the coupling, the gel was washed with PBS and incubated with Minimum Essential Medium (MEM) at \SI{37}{\celsius}. The next day, wild-type MDCK cells were seeded on the gel substrate and given 8 hours to adhere tightly to the substrate. The traction force microscopy was then started. 

The images of the cells and fluorescent beads were monitored using an automated all-in-one microscope with a temperature- and humidity-controlled cell chamber (BZ-X700; Keyence Corporation, Osaka, Japan). We used the 20X (NA = 0.30) Plan Fluor objective lens. Phase-contrast images of the cells and fluorescent images of the beads were captured every 10 min for 2 h. TPA was added to the cultured solution in situ 1 h after the starting time. After the time-lapse acquisition, we detached the cells by adding MEM containing 0.3\% Tween 20. Then, we measured the reference image of the fluorescent beads.

The displacement field of the fluorescent beads was measured by comparing the images with and without cells. The displacement of the beads was calculated using commercial PIV software (Flownizer 2D; Detect Corporation, Tokyo, Japan). For the traction force microscopy, we performed Fourier transform traction cytometry (FTTC) using Matlab software \cite{tanimoto2014}. For the FTTC, we applied the Gaussian filter with a cut-off frequency of \SI{0.7}{\per\micro\meter} to the deformation field.

 A single isolated MDCK cell exerted a traction stress of $F_0 = \SI{400}{\pascal}$ (Fig. \ref{figs3}(a) and (c) blue). The magnitude of ERK activation $\Delta E$ was measured by the ratio of ERK-mCherry translocation into the cell nucleus (Fig. \ref{fig_TPA}(b)). The stress damping began after the ERK activation with \SI{32}{\nano M} TPA and continued for \SI{30}{\minute} \cite{sharma2010}. The traction stress decreased later to $F = \SI{300}{\pascal}$, and the number of traction force spots showing large traction stress was reduced to zero after the addition of TPA (Fig. \ref{figs3}(b) and (c) red). The specificity of the ERK-induced force reduction was also investigated by a MEK inhibitor that suppresses the activated ERK signal (Fig. \ref{figs3}(c) green). The force reduction was not observed with the simultaneous dose of \SI{1}{\micro M} MEK inhibitor and \SI{32}{\nano M} TPA. This result indicates that the reduced traction force results from the activated ERK. We also measured traction force using a microscope under conditions in which two cells were clustered together. The traction force was also decreased in an ERK activity-dependent manner in a small cluster of cells (Fig. \ref{figs3}(d) and (e)). The reduction rate of the traction stress was almost proportional to ERK activation level (Fig. \ref{figs3}(f)).

\begin{figure}[tb]
 \begin{center}
  \includegraphics[width=170mm]{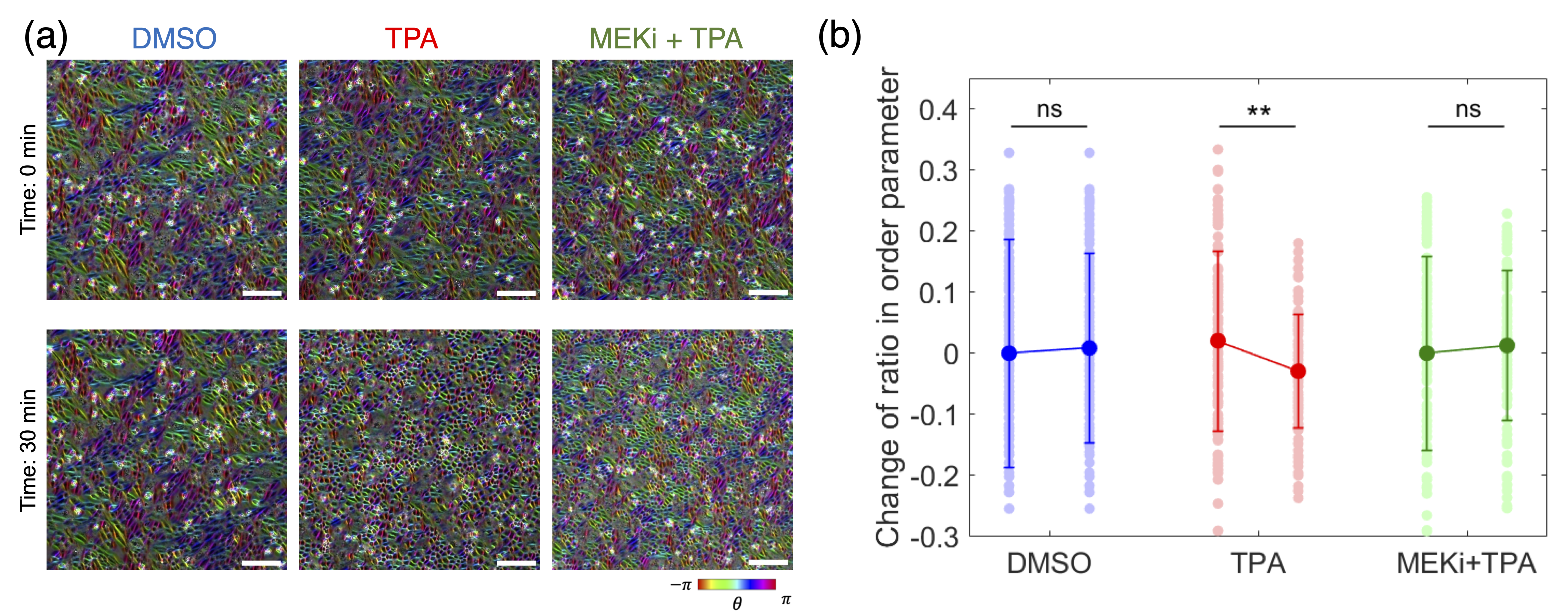}
 \caption{\textbf{ERK-signal dependence in local cell orientation}. (a) Orientation field of MDCK cell monolayer. Control (DMSO, Blue), TPA treated (Red), MEK inhibitor + TPA treated (Green). Scale bars: \SI{50}{\micro\meter}. Images before (time: 0 min) and after (time: 60 min) chemical perturbation are compared. Color code represents the orientation angle $\theta(\bm{x})$. (b) Change of ratio in the orientation order parameter. Error bars indicate SD. Data were analyzed using the Mann-Whitney U-test, ** $p < 0.01$.}\label{fig_force}
 \end{center}
 \end{figure}

\subsection{The effect of adhesive force on the bottom substrate}

\begin{figure}[tb]
 \begin{center}
  \includegraphics[width=120mm]{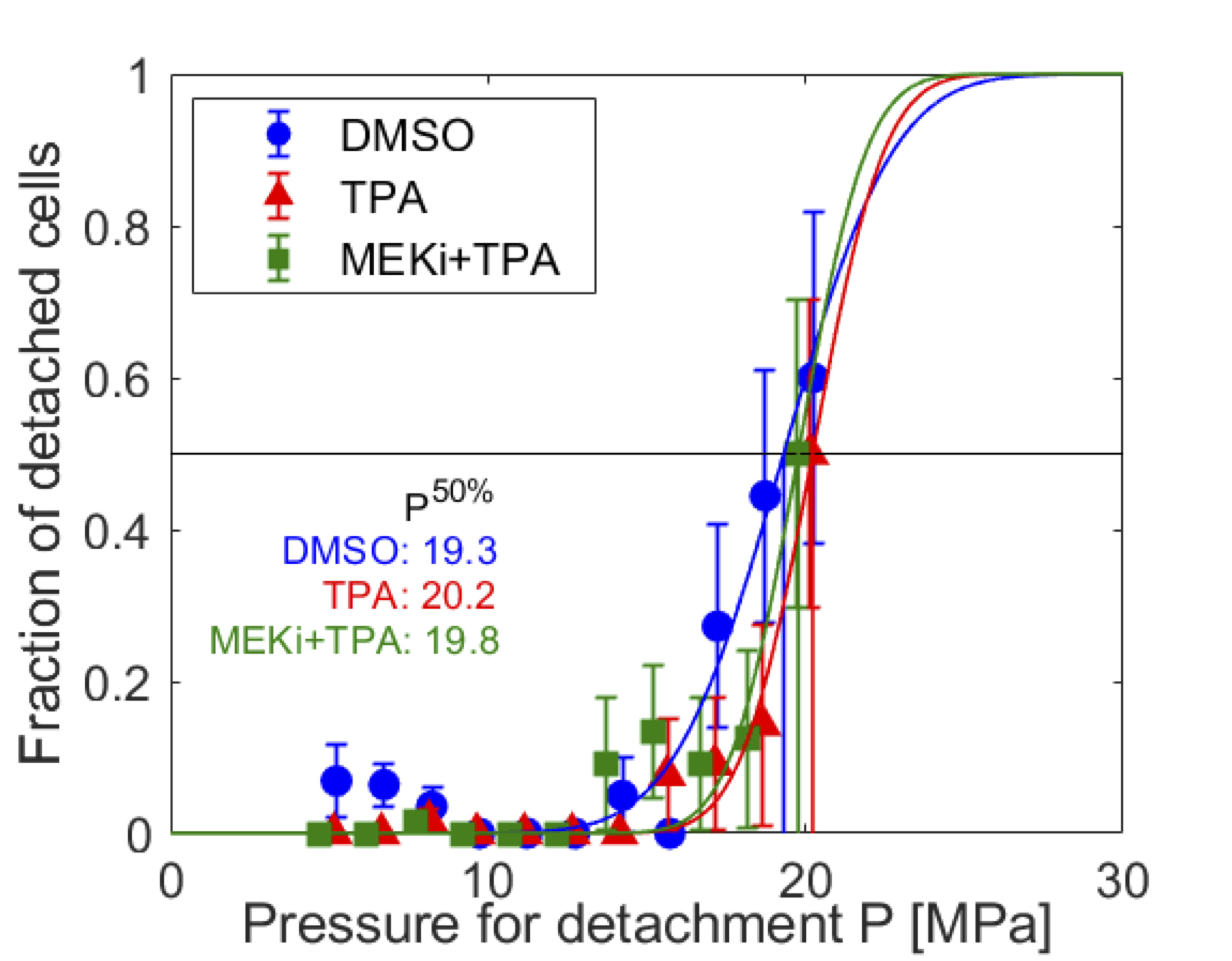}
 \caption{\textbf{The adhesion strength to the substrate}. Pressure intensity required to detach MDCK cells from the substrate surface and percentage of cells detached. Control (DMSO, Blue): 19.3 MPa, TPA treated (Red): 20.2 MPa, MEK inhibitor + TPA treated (Green): 19.8 MPa. \Add{The number of technical replicates is $n=3$ and the number of independent experiments is $N_{exp}=3$. Fitting was performed using the error function, and $P^{50\%}$ is the pressure at which 50\% of the cells in each measurement detach. Error bars indicate SD.}}\label{fig_force}
 \end{center}
 \end{figure}

In this study, we consider the anisotropy of the friction coefficient matrix $\bm{\zeta}=\zeta_0(\bm{I} - \epsilon\bm{Q}$) to be one of important factors for collective migration guided by a signal wave (as shown in Eq.(4) in main text). However, it is also necessary to examine whether the magnitude of adhesion between MDCK cells and the bottom substrate is dependent on ERK signaling or not. To answer this point, we used the shock wave induced by the pulsed laser to detach MDCK cells from their substrates and evaluated detachment pressure, which would be balanced with cell-substrate adhesion stress (Fig. S5). 

To quantify the cell-substrate adhesion strength, we performed the cell detachment assay using the pressure wave generated by a picosecond laser pulse, as described in detail in previous studies \cite{tanaka1,tanaka2,tanaka3,tanaka4}. The fraction of detached cells is plotted as a function of the applied pressure, and the critical pressure $P^{50\%}$ where 50\% of cells were detached was then determined by fitting the plot with an error function.

For TPA-treated MDCK cells (increased ERK activity), their detachment pressure was not significantly different from that of the control group (Fig. S5). Furthermore, no significant difference in detachment was observed when the cells were treated with TPA and a MEK inhibitor at the same time (Fig. S5). These experimental results suggest that the adhesion force between MDCK cells and the base substrate is not dependent on ERK signaling, implying that $\zeta_0$ is not affected by ERK signaling.

\clearpage

\section{Theoretical analysis}
 
\subsection{Continuum model of epithelial cell monolayer with a traveling wave}

We consider the monolayer of epithelial cells to be a viscoelastic sheet adhered to the flat solid substrate in the two-dimensional plane. Because epithelial cells form a thin monolayer, we assume the monolayer of cells is a pseudo-two-dimensional sheet in the $xy$ plane, $\bm{x} = (x,y)$. A cell monolayer is described as a continuum, which has local density $\rho(\bm{x})$ and local velocity $\bm{u}(\bm{x})$. The activity of a signal wave $E(\bm{x})$ controls the internal state of individual cells, and either the activation or deactivation of the signal regulates the contractile force exerted by intracellular actomyosin. The ERK MAP kinase protein is an intracellular protein, and an activated ERK signal propagates in a wave-like manner through the monolayer of cells \cite{AOKI2013,Aoki2017}. Hence, we define the spatial distribution of an activated signal such as ERK MAP kinase by the Gaussian distribution function $\Delta E(\bm{x}) = E(\bm{x})-E_0 = \Delta E_0 \exp(-\frac{\bm{x}^2}{2a^2})$, where $a$ is the typical size of the signal wave. The signal wave propagates along the $x$-axis at a constant velocity, $\bm{u}_l = (u_l,0)$. 

The group of cells moves at a constant velocity, and the balance of forces must be achieved. According to the law of conservation of momentum, the equation of motion with cell migration velocity $\bm{u}(\bm{x}) = (u_x(\bm{x}), u_y(\bm{x}))$ is given by 
\begin{equation}\label{stokeseq2}
\rho \frac{\partial \bm{u}}{\partial t} = - \bm{\zeta} \bm{u} + \bm{\nabla} \cdot \bm{\sigma}  + \eta \nabla^2\bm{u} 
\end{equation}
where $\bm{\zeta}$ is the frictional coefficient matrix per unit area. Cells adhere to each other via cell-cell adhesion junction, and $\bm{\sigma}$ is the internal stress \cite{banerjee2015, yabunaka2017}. The inertia term $d \bm{u} /dt$ is not present due to the sufficiently low Reynolds number because cell migration is slow ($\approx$ few \SI{}{\micro\meter\per\minute}) and the cells are small (few tens of \SI{}{\micro \meter}). In addition, the shear stress between cells $\eta \nabla^2\bm{u}$ is negligible compared to the friction $\bm{\zeta} \bm{u}$ when the cells collectively migrate in one direction. By considering these circumstances with Eq. \eqref{stokeseq2}, we obtain the following force balance equation: 
\begin{equation}\label{forcenalance2}
\bm{\nabla} \cdot \bm{\sigma} = \bm{\zeta}\bm{u}
\end{equation}
\Add{and the internal stress is
\begin{equation}\label{stress}
\bm{\sigma} = -\Pi \bm{I} + k c \bm{I}
\end{equation}
where $\bm{I}$ denotes the identity matrix, the first term is passive stress caused by the pressure $\Pi$, and the second term denotes the signal-dependent stress generated by the molecular motor protein (protein concentration $c$) with a constant $k$.
As presented in the main text, the orientation of the cell monolayer is expressed by the director $\bm{n}(\bm{x}) = (n_x, n_y) = (\cos\theta, \sin\theta)$, with an orientation angle of $\theta(\bm{x})$. The cell population is oriented along $x$-direction in which the signal wave propagates ($\theta \ll$ 1). We define the orientation field tensor as $\bm{Q} = S(\bm{n}\bm{n} - \frac{1}{2}\bm{I})$ where the scalar order parameter $S = \sqrt{\langle \cos2\theta \rangle^2 +\langle \sin 2\theta \rangle^2}$ is determined from the average $\theta$ within the coarse-grained region. $S$ is ERK-signal dependent, i.e. $S = S(E)$. We write the anisotropic friction $\bm{\zeta}$ as
\begin{eqnarray}\label{friction}
\bm{\zeta} &=& \begin{bmatrix}
   \zeta_{xx} & \zeta_{xy} \\
   \zeta_{yx} & \zeta_{yy}
\end{bmatrix} = \zeta_0(\bm{I} - \epsilon \bm{Q}) \nonumber \\ 
&=& \zeta_0 \begin{bmatrix}
   1- (\epsilon S/2)\cos2\theta & (\epsilon S/2)\sin2\theta \\
   (\epsilon S/2)\sin2\theta & 1+(\epsilon S/2)\cos2\theta 
   \end{bmatrix}
\end{eqnarray}
where $\epsilon$ is the scalar friction parameter representing orientational anisotropy \cite{kawaguchi2017}. We can write Eq. S2 as
\begin{equation}\label{balance2x}
\zeta_{xx} u_x + \zeta_{xy}u_y = \frac{\partial}{\partial x}\Bigl( - \Pi + kc \Bigr)
\end{equation} and
\begin{equation}\label{balance2y}
\zeta_{xy} u_x + \zeta_{yy}u_y = \frac{\partial}{\partial y}\Bigl(- \Pi + kc \Bigr)
\end{equation}
where we use $\zeta_{xy}=\zeta_{yx}$. This leads to the following relation
\begin{equation}\label{balance2}
\frac{\partial}{\partial y}(\zeta_{xx} u_x + \zeta_{xy}u_y) -\frac{\partial}{\partial x}(\zeta_{xy} u_x + \zeta_{yy}u_y)=0
\end{equation}
In this study, the timescale for collective migration that we consider is approximately 30 minutes, which is small enough when compared to the time needed for cell division in epithelial cells. We assume that the density of the cell population is conserved and that the equation for continuity holds. Given that the density of the cells in the $xy$ plane is $\rho(\bm{x};E(\bm{x}))$ and their velocity is $\bm{u}(\bm{x})$, the continuum equation is 
\begin{equation}
 \frac{\partial \rho}{\partial t} + \frac{\partial (\rho u_x)}{\partial x}  +\frac{\partial (\rho u_y)}{\partial y} = 0 \label{mass1}
\end{equation}
Furthermore, to investigate collective migration along a traveling wave, it is more convenient to stand in a moving frame coordinate system that travels at the same speed as the signal wave, $\bm{u}_l = (u_l, 0)$, than in a laboratory frame coordinate system. In the coordinate system of the moving frame, the operators of the $x$-coordinate, time derivative, and spatial derivative are rewritten as follows.
\begin{eqnarray}
x' &=& x- u_l t \label{laser_axis1}\\ 
\frac{\partial}{\partial t} &=& - u_l \frac{\partial}{\partial x'}\label{laser_time1}\\ 
\nabla_x &=& \frac{\partial}{\partial x} = \frac{\partial}{\partial x'} \frac{\partial x'}{\partial x} = \frac{\partial}{\partial x'} = \nabla_{x'} \label{laser_axis2} \\ 
\nabla^2 &=& \frac{\partial^2}{\partial x^2} + \frac{\partial^2}{\partial y^2} = \frac{\partial^2}{\partial x'^2} + \frac{\partial^2}{\partial y^2} =\nabla'^2 \label{laser_axis3} \\
\nabla' &=& \left(\frac{\partial}{\partial x'} , \frac{\partial}{\partial y} \right) \label{laser_axis4}
\end{eqnarray}
Since the $y$-coordinate is perpendicular to the ERK wave propagation, there is no change in the differential operator of the $y$-axis. In the above moving frame coordinate system, Eqns. \eqref{balance2} and \eqref{mass1} are rewritten by
\begin{equation}
\frac{\partial}{\partial y}(\zeta_{xx} u_x + \zeta_{xy}u_y) -\frac{\partial}{\partial x'}(\zeta_{xy} u_x + \zeta_{yy}u_y)=0 \label{balance3}
\end{equation}  
and
\begin{equation}
-u_l\frac{\partial \rho}{\partial x'} + \frac{\partial (\rho u_x)}{\partial x'}  +\frac{\partial (\rho u_y)}{\partial y}  = 0. \label{mass2}
\end{equation}
Let $\rho$ and $\zeta_{ij}$ ($i, j=x, y$) be functions of $E(\bm{x})$, and expand them around $E_0$ by $\Delta E_0$ as follows:
\begin{eqnarray}\label{perturbation}
\zeta_{xx} &=& \zeta_{xx0} + \zeta_{xx1} + \cdots \nonumber \\ &=& \zeta_{xx0} (1 + \beta \Delta E) + \cdots \\
\zeta_{yy} &=& \zeta_{yy0} + \zeta_{yy1} + \cdots \nonumber \\ &=& \zeta_{xx0} (1 - \beta \Delta E) + \cdots \\
\zeta_{xy} &=& \zeta_{xy0} + \zeta_{xy1} + \cdots \nonumber\\  &=& \zeta_{xy0} + \zeta_{xx0}\beta' \Delta E + \cdots \\
\rho &=& \rho_0 + \rho_1 + \cdots \nonumber \\ &=& \rho_0(1-\gamma a^2 \nabla^2(\Delta E))+ \cdots 
\end{eqnarray}
where 0, 1, and 2 represent the 0th, 1st, and 2nd order of perturbation, respectively. $\rho_0$ is constant. $\beta = \frac{1}{\zeta_{xx}}\frac{\partial \zeta_{xx}}{\partial E}$, $\beta' = \frac{1}{\zeta_{xx}}\frac{\partial \zeta_{xy}}{\partial E}$, and $\gamma = \frac{\partial \ln\rho}{\partial E}$. 
We consider the simplest terms of $\rho$ that satisfy the mass conservation law and spatial symmetry for any $\Delta E$. $S(E)$ is also comparable to $\Delta E$. We also expand $u_i$ with $\Delta E_0$,
\begin{equation}
u_i = u_{i1}+ u_{i2}+\cdots
\end{equation}
We use $u_{i0}=0$ because the cell population does not move without the ERK signal. 
We then analyze the 1st order terms in $\Delta E_0$ in Eqns. \eqref{balance3} and \eqref{mass2} 
\begin{equation}
\frac{\partial}{\partial y}(\zeta_{xx0} u_{x1} + \zeta_{xy0}u_{y1}) -\frac{\partial}{\partial x'}(\zeta_{xy0} u_{x1} + \zeta_{yy0}u_{y1})=0 \label{balance3_1}
\end{equation}  
and
\begin{equation}
u_l \frac{\partial \rho_1}{\partial x'} = \frac{\partial (\rho_0 u_{x1})}{\partial x'}  +\frac{\partial (\rho_0 u_{y1})}{\partial y}  \label{mass2_1}
\end{equation}
Because $\zeta_{xy0}$ is $\mathcal{O}(S)$ while $\zeta_{xx0}$ and $\zeta_{yy0}$ is $\mathcal{O}(1)$, we neglect the terms that contain $\zeta_{xy0}$ in Eq. \eqref{balance3_1}. In addition, $\zeta_{yy0}=\zeta_{xx0} + O(S)$ but $\mathcal{O}(S)$ is negligible. We then obtain
\begin{equation}
\frac{\partial u_{x1}}{\partial y}  -\frac{\partial u_{y1}}{\partial x'}=0 \label{balance3_2}
\end{equation}  
Furthermore, by using $\rho_1 = - \rho_0 \gamma a^2 \nabla^2 (\Delta E)$ from Eq. (S19), Eq. \eqref{mass2_1} gives
\begin{equation}
 - u_l \gamma a^2 \frac{\partial}{\partial x'} (\nabla^2 \Delta E )= \frac{\partial u_{x1}}{\partial x'}  +\frac{\partial u_{y1}}{\partial y} \label{mass2_2}
\end{equation}  
By summing the partial differentiation of both sides of eq. \eqref{balance3_2} with $y$ and the partial differentiation of both sides of Eq. \eqref{mass2_2} with $x'$, we obtain the following Poisson equation
\begin{equation}
 - u_l \gamma a^2 \frac{\partial^2}{\partial x'^2} (\nabla^2 \Delta E )= \frac{\partial^2 u_{x1}}{\partial x'^2}  +\frac{\partial^2 u_{x1}}{\partial y^2} \label{poisson1}
\end{equation} 
The solution of Eq. \eqref{poisson1} is
\begin{eqnarray}
 u_{x1} &=& - u_l \gamma a^2 \frac{\partial^2 \Delta E}{\partial x'^2} \\
 u_{y1} &=& - u_l \gamma a^2 \frac{\partial^2 \Delta E}{\partial x' \partial y} 
\end{eqnarray} 
The spatial average of $u_{x1}$ within the size of ERK wave reveals the mean speed at the 1st order of $\Delta E_0$ and we define it as
\begin{equation}
\langle u_{x1} \rangle_{x'} = \frac{1}{\sqrt{\pi}a}\int^{+\infty}_{-\infty}dx' u_{x1}
\end{equation}
However, $\langle u_{x1} \rangle_{x'}=0$ (and $\langle u_{y1} \rangle_{x'} = 0$) under the boundary condition of $\Delta E (x=\pm \infty)=\partial_x \Delta E(x=\pm \infty)=0$ (and $\Delta E (y=\pm \infty)=\partial_y \Delta E(y=\pm \infty)=0$). This means that $\bm{u}_1=(u_{x1}, u_{y1})$ do not contribute to net motion under propagating ERK wave.}

$ $
\Add{
Next, we analyze Eq. (S7) and (S8) at the 2nd order of $\Delta E_0$ perturbation. We get the following equation from Eq. (S7):
\begin{equation}\label{balance3_3}
\frac{\partial}{\partial y}(\zeta_{xx1}u_{x1}+\zeta_{xy1}u_{y1}+\zeta_{xx0}u_{x2}+\zeta_{xy0}u_{y2}) - \frac{\partial}{\partial x'}(\zeta_{xy1}u_{x1}+\zeta_{yy1}u_{y1}+\zeta_{xy0}u_{x2}+\zeta_{yy0}u_{y2})=0
\end{equation}
For the coefficients of anisotropic friction, since $\zeta_{xy0}$ is $\mathcal{O}(S)$ but $\zeta_{xx0}$ and $\zeta_{yy0}$ are $\mathcal{O}(1)$, we can neglect $\zeta_{xy0}u_{y2}$ and $\zeta_{xy0}u_x2$ in Eq. \eqref{balance3_3}. In addition, by using $\zeta_{xx1} = - \zeta_{yy1}=\zeta_{xx0}\beta \Delta E$ and $\zeta_{xy1} =\zeta_{xx0}\beta' \Delta E$, Eq. \eqref{balance3_3} leads
\begin{equation}\label{balance3_4}
\frac{\partial}{\partial y}(\beta \Delta E u_{x1} + \beta' \Delta E u_{y1} + u_{x2}) - \frac{\partial}{\partial x'}(\beta' \Delta E u_{x1} - \beta \Delta E u_{y1} + u_{y2})=0
\end{equation}
Moreover, from Eq. (S8), we write the mass conservation as
\begin{eqnarray}\label{mass2_4}
\frac{\partial(\rho_1 u_{x1})}{\partial x'} +\frac{\partial(\rho_1 u_{y1})}{\partial y} &+& \frac{\partial(\rho_0 u_{x2})}{\partial x'} + \frac{\partial(\rho_0 u_{y2})}{\partial y} = 0 \nonumber\\
\frac{\partial u_{x2}}{\partial x'} +\frac{\partial u_{y2}}{\partial y} &=& \gamma a^2\frac{\partial}{\partial x'}(\nabla^2 \Delta E u_{x1}) + \gamma a^2\frac{\partial}{\partial y}(\nabla^2 \Delta E u_{y1})
\end{eqnarray}
Similarly, by summing the partial differentiation of both sides of eq. \eqref{balance3_4} with $y$ and the partial differentiation of both sides of Eq. \eqref{mass2_4} with $x'$, we get:
\begin{eqnarray} 
 \nabla^2 u_{x2} &=& - \frac{\partial^2}{\partial y^2}(\beta \Delta E u_{x1} + \beta' \Delta E u_{y1}) + \frac{\partial^2}{\partial x' \partial y}(\beta' \Delta E u_{x1} - \beta \Delta E u_{y1}) + \gamma a^2\frac{\partial^2}{\partial x'^2}(\nabla^2 \Delta E u_{x1}) +\gamma a^2\frac{\partial^2}{\partial x'\partial y}(\nabla^2 \Delta E u_{y1}) \nonumber \\
  &=& -\nabla^2(\beta \Delta E u_{x1} + \beta' \Delta E u_{y1}) + \frac{\partial^2}{\partial x'^2}(\beta \Delta E u_{x1} + \beta' \Delta E u_{y1}) + \frac{\partial^2}{\partial x' \partial y}(\beta' \Delta E u_{x1} - \beta \Delta E u_{y1}) \nonumber\\ & &+ \gamma a^2\frac{\partial^2}{\partial x'^2}(\nabla^2 \Delta E u_{x1}) +\gamma a^2\frac{\partial^2}{\partial x'\partial y}(\nabla^2 \Delta E u_{y1})  \label{poisson2}
\end{eqnarray} 
We then define the function $U$ as the solution of the following Poisson equation:
\begin{equation}
\nabla^2 U = \frac{\partial}{\partial x'}(\beta \Delta E u_{x1} + \beta' \Delta E u_{y1}) + \frac{\partial}{\partial y}(\beta' \Delta E u_{x1} - \beta \Delta E u_{y1}) + \gamma a^2\frac{\partial}{\partial x'}(\nabla^2 \Delta E u_{x1}) +\gamma a^2\frac{\partial}{\partial y}(\nabla^2 \Delta E u_{y1}) 
\end{equation}
Since all terms on the right-hand side are zero at infinity, $U$ is also zero at infinity, i.e. $U(\pm \infty)=0$. By using Eq. (S33), we can write $u_{x2}$ as
\begin{equation}
u_{x2}=-\beta \Delta E u_{x1} + \beta' \Delta E u_{y1} + \frac{\partial U}{\partial x'}
\end{equation}
The spatial average of $u_{x2}$ in $x$ is
\begin{eqnarray}
    \langle u_{x2} \rangle_{x'} &=& \frac{1}{\sqrt{\pi}a}\int^{+\infty}_{-\infty}dx'(-\beta \Delta E u_{x1} + \beta' \Delta E u_{y1}) +  \frac{1}{\sqrt{\pi}a}U|^{+\infty}_{-\infty} \nonumber \\
    &=& \frac{1}{\sqrt{\pi}a}\int^{+\infty}_{-\infty}dx'(-\beta \Delta E u_{x1} + \beta' \Delta E u_{y1}). 
\end{eqnarray}
In the integral calculation of Eq. (S35), we considered whether $\beta' \Delta E u_{y1}$ could be negligible or not. By using $u_{y1} = -u_l \gamma a^2 \frac{\partial}{\partial x'}\frac{\partial}{\partial y}\Delta E$, we can write the second term on the right-hand side of Eq. (S35) by
\begin{equation}
\frac{1}{\sqrt{\pi}a}\int^{+\infty}_{-\infty}dx' \beta' \Delta E u_{y1}=
-\frac{\beta'u_l \gamma a}{\sqrt{\pi}}\int^{+\infty}_{-\infty}dx' \Delta E \frac{\partial}{\partial x'}\frac{\partial}{\partial y}\Delta E.
\end{equation}
Suppose that there is no diagonal deformation in the $xy$ plane and the Gaussian shape is retained in the $y$-axis direction. We now consider a variable separation of $\Delta E$ into an $x$-dependent function and a $y$-dependent function as $\Delta E(x,y) = A(x)B(y)$ with $A(\pm \infty)=\partial A/\partial x' (\pm\infty)=0$. In this case, 
\begin{eqnarray}
-\frac{\beta'u_l \gamma a}{\sqrt{\pi}}\int^{+\infty}_{-\infty}dx' \Delta E \frac{\partial}{\partial x'}\frac{\partial}{\partial y}\Delta E &=& -\frac{\beta'u_l \gamma a}{\sqrt{\pi}}\int^{+\infty}_{-\infty}dx' AB \frac{\partial A}{\partial x'}\frac{\partial B}{\partial y} \nonumber \\
 &=& -\frac{\beta'u_l \gamma a}{\sqrt{\pi}}B\frac{\partial B}{\partial y}\int^{+\infty}_{-\infty}dx' A \frac{\partial A}{\partial x'} \nonumber \\
  &=& -\frac{\beta'u_l \gamma a}{2\sqrt{\pi}}B\frac{\partial B}{\partial y}A^2|^{+\infty}_{-\infty}=0 
\end{eqnarray}
meaning that the second term in the integral in Eq. (S35) does not contribute to the net motion. Therefore, we can write $\langle u_{x2} \rangle_{x'}$ as
\begin{eqnarray}
    \langle u_{x2} \rangle_{x'} &=& \frac{1}{\sqrt{\pi}a}\int^{+\infty}_{-\infty}dx'(-\beta \Delta E u_{x1}) = \frac{\beta \gamma u_l a}{\sqrt{\pi}}\int^{+\infty}_{-\infty}dx'\Bigl(\Delta E \frac{\partial^2 \Delta E}{\partial x'^2}\Bigr) \nonumber \\
     &=& \frac{\beta \gamma u_l a}{\sqrt{\pi}}\Biggl[\Delta E \frac{\partial^2 \Delta E}{\partial x'^2} \Bigl|^{+\infty}_{-\infty} - \int^{+\infty}_{-\infty}dx'\Bigl(\frac{\partial \Delta E}{\partial x'} \Bigr)^2\Biggr] \nonumber \\
     &=& -\frac{\beta \gamma u_l a}{\sqrt{\pi}}\int^{+\infty}_{-\infty}dx'\Bigl(\frac{\partial \Delta E}{\partial x'} \Bigr)^2 
\end{eqnarray}
This solution of $\langle u_{x2} \rangle_{x'}$ is given in Eq. (24) in the main text. If ERK signal distribution has a point satisfying $\frac{\partial \Delta E}{\partial x'} \neq 0$, $ \langle u_{x2} \rangle_{x'} \neq 0$. The final expression for $\langle u_{x2} \rangle_{x'}$ is
\begin{equation}
    \langle u_{x2} \rangle_{x'} = -\beta \gamma u_l a^2 \Bigl\langle\Bigl(\frac{\partial \Delta E}{\partial x'} \Bigr)^2 \Bigr\rangle_{x'}
\end{equation}
which means that the net motion occurs in $x$ direction.}
\Add{
Furthermore, we can apply the same calculation by subtracting the partial differentiation of both sides of Eq. \eqref{balance3_4} with $x'$ and the partial differentiation of both sides of Eq. \eqref{mass2_4} with $y$.
\begin{eqnarray} 
 \nabla^2 u_{y2} &=& \frac{\partial^2}{\partial x' \partial y}(\beta \Delta E u_{x1} + \beta' \Delta E u_{y1}) - \frac{\partial^2}{\partial x'^2}(\beta' \Delta E u_{x1} - \beta \Delta E u_{y1}) + \gamma a^2\frac{\partial^2}{\partial x' \partial y}(\nabla^2 \Delta E u_{x1}) + \gamma a^2\frac{\partial^2}{\partial y^2}(\nabla^2 \Delta E u_{y1}) \nonumber \\
  &=& \frac{\partial^2}{\partial x' \partial y}(\beta \Delta E u_{x1} + \beta' \Delta E u_{y1}) - \frac{\partial^2}{\partial x'^2}(\beta' \Delta E u_{x1} - \beta \Delta E u_{y1}) \nonumber \\ & & + \gamma a^2\frac{\partial^2}{\partial x' \partial y}(\nabla^2 \Delta E u_{x1}) - \gamma a^2\frac{\partial^2}{\partial x'^2}(\nabla^2 \Delta E u_{y1}) +\gamma a^2 \nabla^2(\nabla^2 \Delta E u_{y1}) \label{poisson3}
\end{eqnarray} 
Instead of $U$, we also define the function $V$ as the solution of the following Poisson equation:
\begin{equation}
\nabla^2 V = \frac{\partial}{\partial y}(\beta \Delta E u_{x1} + \beta' \Delta E u_{y1}) - \frac{\partial}{\partial x'}(\beta' \Delta E u_{x1} - \beta \Delta E u_{y1}) + \gamma a^2\frac{\partial}{\partial y}(\nabla^2 \Delta E u_{x1}) - \gamma a^2\frac{\partial}{\partial x'}(\nabla^2 \Delta E u_{y1})
\end{equation}
We write $u_{y2}$ by using $V$ as
\begin{equation}
u_{y2} = \gamma a^2 \nabla^2 \Delta E u_{y1} + \frac{\partial V}{\partial x'}
\end{equation}
The spatial average of $u_{y2}$ within the size of ERK wave is given by
\begin{eqnarray}
\langle u_{y2} \rangle_{x'} &=& \frac{1}{\sqrt{\pi}a} \int^{+\infty}_{-\infty}dx'\gamma a^2 \nabla^2 \Delta E u_{y1} +  \frac{1}{\sqrt{\pi}a} V|^{+\infty}_{-\infty} \nonumber \\
&=& \frac{1}{\sqrt{\pi}a} \int^{+\infty}_{-\infty}dx'\gamma a^2 \nabla^2 \Delta E u_{y1}.
\end{eqnarray}
By using $\Delta E(x,y) = A(x)B(y)$ with $A(\pm \infty)=\partial A/\partial x' (\pm\infty)=0$), 
\begin{eqnarray}
\langle u_{y2} \rangle_{x'} &=& \frac{1}{\sqrt{\pi}a} \int^{+\infty}_{-\infty}dx'\gamma a^2 \nabla^2 \Delta E u_{y1} = -\frac{u_l \gamma^2 a^3}{\sqrt{\pi}}\int^{+\infty}_{-\infty}dx' \nabla^2 \Delta E \frac{\partial}{\partial x'}\frac{\partial}{\partial y}\Delta E \\ &=& -\frac{u_l \gamma^2 a^3}{\sqrt{\pi}}\int^{+\infty}_{-\infty}dx' \Bigl(B \frac{\partial^2 A}{\partial x'^2}+A\frac{\partial^2 B}{\partial y^2}\Bigr)\frac{\partial A}{\partial x'}\frac{\partial B}{\partial y} \nonumber \\
 &=& -\frac{u_l \gamma^2 a^3}{\sqrt{\pi}}\Bigl(B\frac{\partial B}{\partial y}\int^{+\infty}_{-\infty}dx' \frac{\partial^2 A}{\partial x'^2}\frac{\partial A}{\partial x'}+\frac{\partial B}{\partial y}\frac{\partial^2 B}{\partial y^2}\int^{+\infty}_{-\infty}dx' A\frac{\partial A}{\partial x'}\Bigr) \nonumber \\
  &=& -\frac{u_l \gamma^2 a^3}{2\sqrt{\pi}}\Biggl(B\frac{\partial B}{\partial y}\Bigl[\Bigl(\frac{\partial A}{\partial x'}\Bigr)^2\Bigr]^{+\infty}_{-\infty}+\frac{\partial B}{\partial y}\frac{\partial^2 B}{\partial y^2}[A^2]^{+\infty}_{-\infty}\Biggr)=0 
\end{eqnarray}
meaning that the integrated function in Eq. (S43) becomes zero after spatial averaging. Then, the final expression of $\langle u_{y2} \rangle_{x'}$ is
\begin{equation}
\langle u_{y2} \rangle_{x'} =0, 
\end{equation}
which indicates that net motion in $y$ direction does not occur. Therefore, from Eqs. (S37) and (S46), the mean speed of collective migration $\langle\bm{u}_2\rangle_{x'}=(\langle u_{x2}\rangle_{x'}, 0)$ at $y=0$ is 
\begin{equation}
\langle\bm{u}_2\rangle_{x'}= - \bm{u}_l\frac{\beta\gamma}{2}(\Delta E_0)^2
\end{equation}
}

\subsection{Mechanical deformation and contractile force}

We present the relationship between the contractile force among cells and the signal wave. Aoki et al. have shown that ERK MAP kinase is involved in the phosphorylation of the myosin light chain in cells and that it localizes myosin to the posterior end of the cell \cite{Aoki2017}, suggesting that ERK activity regulates the contractile force of myosin. Localized signal activity increases the contractile force, causing a pressure gradient between cells, and the gradient of contractile force results in changes in the cell density.

To investigate the dynamics of cell density $\rho(\bm{x})$ driven by the interplay of the contractile stress $\sigma$ and signal activity $E(\bm{x})$, the Voigt model for cell deformation is considered \cite{banerjee2015}. The density of the cell monolayer should be uniform and constant $\rho_0$ in the steady state. The strain $\varepsilon$ is expressed by the relative change in cell density:  $\varepsilon = \Delta \rho/\rho_0 = (\rho-\rho_0)/\rho_0$. The strain dynamics can be described as
\begin{equation}
\frac{\partial \varepsilon}{\partial t} + \nabla \cdot (\varepsilon \bm{u}) = -\frac{1}{\tau} \varepsilon  + \frac{\sigma}{\eta_c},
\end{equation} 
where $\tau=\eta_c / G$ is the relaxation time for the strain, $\eta_c$ is the viscosity of the cell monolayer (cytosolic viscosity), $G$ is the elastic modulus, and $\sigma = \sigma(E)$ is the activity-dependent contractile stress. For simplicity, we assume that elastic restoration is more dominant than viscous relaxation at the timescale of cell motility and that the viscous damping $\eta (\partial \Delta l/ \partial t)$ can be neglected in this model. Thus, the strain $\varepsilon$ induced by the contractile force is given by 
\begin{equation}
\varepsilon = \frac{\sigma}{G}. \label{cellvoigt1}
\end{equation}

The change in density can also be approximated as the change in characteristic length $\Delta l$ ($l$ is the length of a cell along the propagating signal wave). The change in density is thus expressed as an inversely proportional relationship: $\Delta \rho \propto 1/\Delta l$. This simplification allows us to rewrite the strain as $\varepsilon = \Delta \rho(\bm{x};E(\bm{x}))/\rho_0 \propto -\Delta l/l_0 $, with $\Delta l / l_0 \ll 1$. Then, the relative change from the original density $\rho_0$ is given by 
\begin{equation}
\frac{\Delta \rho(\bm{x};E(\bm{x}))}{\rho_0} = \frac{\sigma}{G}. \label{cellvoigt2}
\end{equation}
These equations show that the contractile force is linearly related to the density change under the signal wave.

\subsection{The effect of viscoelastic deformation} 
Epithelial cells exhibit viscoelastic deformation, not only elastic deformation and recovery. This rheological property is particularly important in explaining the frequency dependence of the collective migration when the velocity of the ERK wave is altered, as seen in Fig. 4 in the main text. Therefore, in this section, the deformation of the cell density is re-analyzed with the viscoelastic model.

In the previous section, the density change was regarded as time-invariant and relaxation processes negligible. However, cells undergo viscoelastic deformation with the relaxation time $\tau_d=\eta_c/G$ and the viscosity of the intracellular space $\eta_c$. In this section, we derive the generalized form of the coefficient for the cell density $\gamma$ by considering the viscoelastic deformation due to the signal wave.

The viscoelastic change in cell density $\rho$ is given by the following Voigt model
\begin{equation}
\frac{d  \Delta\rho}{d t}= -\frac{1}{\tau_d} \Delta \rho + \frac{\rho \sigma}{\eta_c}, \label{relax1}
\end{equation}
where $\Delta\rho = \rho - \rho_0$ and $\rho_0$ is the density at the ground state of the ERK activity ($\Delta E = 0$). Given that the maximum value of the density change $\Delta \rho_{max}$ when the stress $\sigma$ works at $t > 0$ is \Add{$-\rho_0 \gamma a^2 \nabla^2 \Delta E$}, solving Eq. \eqref{relax1} yields
\Add{
\begin{equation}
\frac{\Delta \rho(t)}{\rho_0} = - \gamma a^2 \left [ 1- {\rm exp} \left ( -\frac{t}{\tau_d} \right) \right]\nabla^2 \Delta E. \label{relax2}
\end{equation}
}
The time $t$ at which the ERK activates and stress works can be determined by using the wavelength $\lambda$ and velocity $u_l$ of the ERK wave, $t=\lambda/u_l$. The density change caused by the repeated deformation and relaxation can be approximated by
\Add{
\begin{equation}
\frac{\Delta \rho(\lambda /u_l)}{\rho_0} = - \gamma a^2 \left [ 1- {\rm exp} \left ( -\frac{\lambda}{u_l \tau_d} \right) \right]\nabla^2 \Delta E. \label{relax3}
\end{equation}
}
Since $\gamma$ in Eqs. (21) to (43) in the main test can be replaced with $\Gamma_{\tau_d} = \gamma  \left [ 1- {\rm exp} \left ( - \lambda/u_l \tau_d \right) \right]$, we obtain an extended form of collective migration as
\begin{equation}
\bm{u}_{mig} = - \bm{u}_l \frac{\beta \Gamma_{\tau_d}}{2} (\Delta E_0)^2, \label{migvel2}
\end{equation}
which indicates that the viscoelastic deformation of the cells induced by the ERK activity thus modifies $\gamma$ in migration speed. 

\subsection{Wave-speed dependence and optimal migration velocity}

Eq.\eqref{migvel2} shows the extended form of collective migration with the viscoelastic coefficient $\Gamma_{\tau_d} = \gamma \left[1- \exp \left (-\frac{\lambda}{u_l\tau_d}\right)\right]$,
\begin{equation}
\bm{u}_{mig} = - \bm{u}_l \frac{\beta \gamma}{2}  \left[1- {\rm exp} \left(- \frac{\lambda}{u_l \tau_d} \right) \right] \left( \Delta E_0 \right)^2. \label{migvel3}
\end{equation}
The exponential term ${\rm exp}(-\lambda/u_l \tau_d)$ is less than $1$, so the difference $1-{\rm exp}(-\lambda/u_l \tau_d)$ must be positive. This means that the viscoelastic relaxation does not affect the direction of the collective migration, i.e., the migration occurs opposite to the ERK wave. However, the viscoelasticity does change the speed of the collective migration. Because ${\rm exp}(-\lambda/u_l \tau_d)$ goes to $1$ as $\tau_d$ increases, in the limit of infinite $\tau_d$, $1-{\rm exp}(-\lambda/u_l \tau_d)$ tends to zero, and therefore, $\bm{u}_{mig}$ also tends to zero. 

To clarify the dependence of the migration velocity on the relaxation constant $\tau_d$, we analyze the migration speed under two extreme conditions of slow and fast ERK waves. We note that the frequency $\frac{u_l}{\lambda}$ of ERK wave activation, which the cell receives periodically, can change depending on the velocity of ERK wave $u_l$. We first consider the slow ERK wave with $\frac{\lambda}{u_l \tau_d} \gg 1$. The exponential term in $\Gamma_{\tau_d}$ is approximated as ${\rm exp}(-\frac{\lambda}{u_l \tau_d}) \approx 0$, and the migration velocity is given by
\begin{equation}
\bm{u}_{mig} = - \bm{u}_l \frac{ \beta \gamma}{2} (\Delta E_0)^2. \label{migvel4}
\end{equation}
This means that the migration speed $|\bm{u}_{mig}|$ increases proportionally to $\bm{u}_l$ within the regime of a slow ERK wave. In the experiment, the migration speed exhibits a linear increase with a linear increase in the speed of a slow ERK wave, which is in agreement with Eq. \eqref{migvel4}.

Next, we consider the effect of viscoelastic relaxation on the collective migration under a fast ERK wave. If the ERK activation occurs at a higher frequency $\frac{\lambda}{u_l \tau_d} \ll 1$, multiple ERK waves can be propagated during the relaxation period. Under this limit of a fast signal wave, the exponential term of $\Gamma_{\tau_d}$ can be expanded as
\begin{equation}
 {\rm exp} \left(- \frac{\lambda}{u_l \tau_d} \right)= 1- \frac{\lambda}{u_l \tau_d} + O\left(\left(\frac{\lambda}{u_l\tau_d} \right)^2 \right).
\end{equation}
Due to this approximation, the $\Gamma_{\tau_d}$ of viscoelastic deformation is inversely proportional to the velocity of the ERK wave. On the other hand, the migration speed is proportional to the speed of the ERK wave, and it cancels out the effect of the reduced viscoelastic deformation. Hence, the velocity of collective cell migration plateaus due to viscoelasticity: 
\begin{equation}
u_{mig} = - \frac{\lambda}{2} \frac{1}{\tau_d} \beta \gamma \left(\Delta E_0 \right)^2. \label{migvel5}
\end{equation}
However, the experimental data in Fig. 3 in the main text shows that the migration speed starts to decrease around $u_l \sim \SI{2.0}{\micro\meter\per\minute}$. To explain the wave-speed dependence at higher $u_l$, we need to consider additional effects from the relaxation of the ERK activity.

The relaxation time of signal activity should be considered to explain the decay of migration velocity for fast ERK waves. An activated ERK signal returns to the ground state $E_0$ at the relaxation time $\tau_E$ \cite{AOKI2013}. For the signaling pathway of ERK MAP kinase, activated ERK signals are known to increase the ERK activity of neighboring cells. Thus, ERK activity is transmitted through cell-to-cell interactions. This signal transmission in space is regarded as diffusion. Considering these temporal and spatial relaxations, the dynamics of signal activity can be expressed as
\begin{equation}
\frac{\partial \Delta E}{\partial t} = -\frac{\Delta E}{\tau_E} + D\nabla^2 (\Delta E)+ I_s, \label{erk1}
\end{equation}
where $D$ is the effective diffusion coefficient of the signal activity, and $I_s = I_s(x-u_l t, 0)$ is the point source of the signal wave. The diffusion term on the right side of Eq. (\ref{erk1}) represents the isotropic propagation of the signal activity in space. In experiments, we suppressed the intercellular transmission of the ERK activity by TNF alpha processing inhibitor (TAPI-1). By neglecting the diffusion term, the relaxation dynamics of the signal activity can be rewritten as 
\begin{equation}
\frac{d \Delta E}{d t}=-\frac{\Delta E}{\tau_E} + I_s. \label{erk2}
\end{equation}

To calculate the steady signal activity on the moving frame, the Fourier transform of Eq. (\ref{erk2}) was performed: 
\begin{equation}
E(\bm{x}')=\frac{1}{2 \pi} \int_{-\infty}^{\infty} d \bm{k} \hat{E}(\bm{k}) e^{-i \bm{k} \cdot \bm{x}'},
\end{equation}
where the ERK activity in Fourier space is $\hat{E}(\bm{k})$ with wave number $\bm{k}$. \Add{Suppose that the point source is $I_s = I_0 {\rm exp} (-(x'^2+y^2)/2a^2)$} with width $a$, Eq. (\ref{erk2}) in the moving frame leads
\begin{equation}
\hat{E}(\bm{k})= \frac{I_0}{(1/\tau_E)-i\bm{u}_l \cdot \bm{k}} {\rm exp} \left(-\frac{a^2 k^2}{2} \right). \label{erk3}
\end{equation}
By applying the inverse Fourier transform of Eq. \eqref{erk3}, the steady-state distribution of the signal activity can be obtained in the moving frame as
\begin{eqnarray}
E(\bm{x}') &=& \int^{+\infty}_{-\infty}\int^{+\infty}_{-\infty}dk_{x'}dk_y\frac{I_0}{(1/\tau_E)-iu_l k_{x'}} {\rm exp} \left(-\frac{a^2 k^2}{2} \right){\rm exp}(-ik_{x'}x'-ik_yy) \nonumber \\
&=& \int^{+\infty}_{-\infty} dk_y{\rm exp} \left(-\frac{a^2 k_y^2}{2} \right){\rm exp}(-ik_yy) \int^{+\infty}_{-\infty}dk_{x'}\frac{I_0}{(1/\tau_E)-iu_l k_{x'}} {\rm exp} \left(-\frac{a^2 k_{x'}^2}{2} \right){\rm exp}(-ik_{x'}x'). \label{erk5}
\end{eqnarray}

\Add{When the propagation speed of ERK signals is fast, the distribution of ERK activity becomes a stretched distribution that extends in the $x$-axis direction and deviates from the Gaussian distribution along the $x$-axis. However, the shape of ERK signal distribution in $y$-axis is retained in Gaussian shape. In this case, we can do variable separation as we did in Eq. (S37) for $\Delta E(x',y)$, so the distortion in the $x$-axis does not change Eqs. (S39),(S45),(S46) as long as the boundary conditions at infinity are satisfied.}

\subsection{Collective cell migration in wound healing}

We extend the theoretical model of the velocity of collective migration by the signal wave to the system of wound healing. Two distinct collective migrations are involved in the process of wound healing. The first is the collective movement driven by the signal wave. The second is the motion of both leader cells and follower cells at the boundary of the wound. Hence, collective migration in wound healing can be expressed as the sum of these two modes. 

During wound healing, the boundary of the cell monolayer moves as the wound region is filled by the leading cells. Given this boundary motion as flux $\bm{J}_{w}$, the mass conservation (continuum equation) in wound healing is 
\begin{equation}
\frac{\partial \rho}{\partial t} + \nabla \cdot (\rho \bm{u}) = - \nabla \cdot \bm{J}_{w}. \label{wound1} 
\end{equation}
The velocity of the moving front (in the wound region) is $\bm{u}_w = (u_w, 0)$, and the flux of the moving front is given by $\bm{J}_{w} = \rho \bm{u}_w $. \Add{By using Eqs. (S9) – (S13), the modified continuum equation for the moving frame is obtained.
 \begin{equation}
-(u_l - u_w)\frac{\partial \rho}{\partial x'} + \frac{\partial (\rho u_x)}{\partial x'} + \frac{\partial (\rho u_y)}{\partial y}=0. \label{wound2} 
\end{equation}
Here, we assume that $u_w$ is constant. In this case, the force balance equation Eq. (S14) is unchanged. Thus, all calculations in section II A hold only when we change $u_l$ to $u_l-u_w$. Eq. (S65) means that the relative velocity of the ERK wave can be considered to be  $u_l-u_w$, since the ERK wave propagates in the opposite direction to the leading edge of cell monolayer moving at the speed of $u_w$.} Finally, the velocity of collective cell migration is derived as
\begin{equation}
\bm{u}_{mig}^{wh} = - \frac{\bm{u}_l-\bm{u}_w}{2} \beta \Gamma_{\tau_d} \left( \Delta E_0 \right)^2, \label{mig_wound}
\end{equation}
\Add{where $\bm{u}_{mig}^{wh}=(\langle u_{x2}\rangle_{x},\langle u_{y2}\rangle_{x})$.}

\bibliography{erk}

\end{document}